\documentclass[a4paper,11pt]{article}
\pdfoutput=1 

\usepackage{jheppub} 

\usepackage[T1]{fontenc} 
\usepackage{comment}

\title{\boldmath  One loop photon-graviton mixing in an electromagnetic field: Part 3}

\author[a]{N. Ahmadiniaz,}
\author[b]{F. Bastianelli,}
\author[c]{F. Karbstein}
\author[d,1]{and C. Schubert \note{Corresponding author}}

\affiliation[a]{Helmholtz-Zentrum Dresden-Rossendorf, Bautzner Landstra\ss e 400, 01328 Dresden, Germany}
\affiliation[b]{Dipartimento di Fisica ed Astronomia, Universit\`a di Bologna, Via Irnerio 46, I-40126 Bologna, Italy
and INFN, Sezione di Bologna, Via Irnerio 46, I-40126 Bologna, Italy}
\affiliation[c]{Helmholtz-Institut Jena, Fr\"obelstieg 3, 07743 Jena, Germany and Theoretisch-Physikalisches Institut, Friedrich-Schiller-Universit\"at Jena, Max-Wien-Platz 1, 07743 Jena, Germany}
\affiliation[d]{Facultad de Ciencias Físico-Matemáticas, Universidad Michoacana de San Nicolás de Hidalgo, Avenida Francisco J. Mújica, 58060 Morelia, Michoacán, Mexico}

\emailAdd{n.ahmadiniaz@hzdr.de}
\emailAdd{bastianelli@bo.infn.it}
\emailAdd{felix.karbstein@uni-jena.de}
\emailAdd{christianschubert137@gmail.com}

\abstract{Photon-graviton conversion in an electromagnetic field is a well-known prediction of Einstein-Maxwell theory. First discussed at tree-level by Gertsenshtein in 1962,
more recently it has been shown to lead to magnetic dichroism starting from one-loop. While previously only two diagrams were assumed to contribute to 
this one-loop photon-graviton amplitude in a constant electromagnetic field, here we point out the existence of a third one involving a tadpole subdiagram. As shown by H. Gies and one of the authors
 in 2016 for the pure QED case, such diagrams cannot be omitted in general even though the tadpole formally vanishes. 
After a short review of the calculation of one-loop photon-graviton amplitudes in the worldline formalism, we use this formalism
for a unified calculation of all three diagrams. Although phenomenologically this amplitude is mainly of interest for the case of the spinor loop in a magnetic field,
here we will also include the scalar loop and the electric field component, since the computational effort is essentially the same. 
We show that the tadpole diagram, although contributing to the amplitude, does not contribute to the magnetic dichroism. 
The gravitational Ward identity provides a useful check.
}

\begin{document} 
\maketitle
\flushbottom

\newcommand{\cP}{\cal P}
\newcommand{\cG}{{\cal G}}
\newcommand{\pol}{\varepsilon}
\newcommand{\cD}{\cal D}
\newcommand{\cZ}{{\cal Z}}
\newcommand{\cS}{{\cal S}}
\newcommand{\cA}{{\cal A}}

\newcommand{\bB}{\bf B}
\newcommand{\bE}{\bf E}

\def\veps{\varepsilon}
\def\veps#1{\varepsilon_{#1}}
\def\intT{{\dps\int_{0}^{\infty}} \!\!\! dT\, {\rm e}^{-m^2T}}
\def\eTx{{\rm e}^{-\frac{1}{4}\int_0^Td\tau\dot{x}^2}}
\def\eTq{{\rm e}^{-\frac{1}{4}\int_0^Td\tau\dot{q}^2}}

\def\PITD{{(4\pi T)}^{-{D\over 2}}}

\def\detsin{{\rm det}^{\frac{1}{2}} \biggl[ \frac{\cZ}{\sin \cZ}\biggr]}
\def\dettan{{\rm det}^{\frac{1}{2}} \biggl[ \frac{\cZ}{\tan \cZ}\biggr]}

\def\Nint{\prod_{i=1}^N\int_0^T \!\!\! d\tau_i}

\def\eqa{\!\!&=&\!\!}
\def\ccr{\nonumber\\}

\def\la{\langle}
\def\ra{\rangle}

\def\del{\Delta}
\def\ddel{{}^\bullet\! \Delta}
\def\deld{\Delta^{\hskip -.5mm \bullet}}
\def\ddeld{{}^{\bullet}\! \Delta^{\hskip -.5mm \bullet}}
\def\dddel{{}^{\bullet \bullet} \! \Delta}

\newcommand{\ba}{\begin{array}}
\newcommand{\ea}{\end{array}}
\newcommand{\identy}{1\!\!1}
\def\ni{\noindent}
\def\la{\langle}
\def\ra{\rangle}

\def\nonu{\nonumber}

\def\exmn{\Bigl(\mu \leftrightarrow \nu \Bigr)}

\def\rld{\rlap{\,/}D}
\def\rldd{\rlap{\,/}\nabla}
%
\def\half{{1\over 2}}
\def\third{{1\over3}}
\def\fourth{{1\over4}}
\def\fifth{{1\over5}}
\def\sixth{{1\over6}}
\def\seventh{{1\over7}}
\def\eigth{{1\over8}}
\def\ninth{{1\over9}}
\def\tenth{{1\over10}}
\def\bN{\mathop{\bf N}}
\def\R{{\rm I\!R}}
\def\Eins{{\mathchoice {\rm 1\mskip-4mu l} {\rm 1\mskip-4mu l}
{\rm 1\mskip-4.5mu l} {\rm 1\mskip-5mu l}}}
\def\Z{{\mathchoice {\hbox{$\sf\textstyle Z\kern-0.4em Z$}}
{\hbox{$\sf\textstyle Z\kern-0.4em Z$}}
{\hbox{$\sf\scriptstyle Z\kern-0.3em Z$}}
{\hbox{$\sf\scriptscriptstyle Z\kern-0.2em Z$}}}}
\def\abs#1{\left| #1\right|}
\def\com#1#2{
        \left[#1, #2\right]}
\def\square{\kern1pt\vbox{\hrule height 1.2pt\hbox{\vrule width 1.2pt
   \hskip 3pt\vbox{\vskip 6pt}\hskip 3pt\vrule width 0.6pt}
   \hrule height 0.6pt}\kern1pt}
      \def\boxop{{\raise-.25ex\hbox{\square}}}
\def\contract{\makebox[1.2em][c]{
        \mbox{\rule{.6em}{.01truein}\rule{.01truein}{.6em}}}}
\def\ltap{\ \raisebox{-.4ex}{\rlap{$\sim$}} \raisebox{.4ex}{$<$}\ }
\def\gtap{\ \raisebox{-.4ex}{\rlap{$\sim$}} \raisebox{.4ex}{$>$}\ }
\def\mn{{\mu\nu}}
\def\rs{{\rho\sigma}}
\newcommand{\Det}{{\rm Det}}
\def\Tr{{\rm Tr}\,}
\def\tr{{\rm tr}\,}
\def\sumij{\sum_{i<j}}
\def\e{\,{\rm e}}
\def\pa{\partial}
\def\dA{\partial^2}
\def\ddx{{d\over dx}}
\def\ddt{{d\over dt}}
\def\der#1#2{{d #1\over d#2}}
\def\lie{\hbox{\it \$}} 
\def\partder#1#2{{\partial #1\over\partial #2}}
\def\secder#1#2#3{{\partial^2 #1\over\partial #2 \partial #3}}
%
\newcommand{\be}{\begin{equation}}
\newcommand{\ee}{\end{equation}\noindent}
\newcommand{\bear}{\begin{eqnarray}}
\newcommand{\ear}{\end{eqnarray}\noindent}
\newcommand{\benn}{\begin{enumerate}}
\newcommand{\enn}{\end{enumerate}}
\newcommand{\veject}{\vfill\eject}
\newcommand{\ven}{\vfill\eject\noindent}
%
\def\eq#1{{eq. (\ref{#1})}}
\def\eqs#1#2{{eqs. (\ref{#1}) -- (\ref{#2})}}
%
\def\totint{\int_{-\infty}^{\infty}}
\def\posint{\int_0^{\infty}}
\def\negint{\int_{-\infty}^0}
\def\pint{{\dps\int}{dp_i\over {(2\pi)}^d}}
%
\newcommand{\GeV}{\mbox{GeV}}
\def\FFdual{F\cdot\tilde F}
\def\bra#1{\langle #1 |}
\def\ket#1{| #1 \rangle}
\def\braket#1#2{\langle {#1} \mid {#2} \rangle}
\def\vev#1{\langle #1 \rangle}
\def\rightvac{\mid 0\rangle}
\def\leftvac{\langle 0\mid}
\def\ihbar{{i\over\hbar}}
\def\slash#1{#1\!\!\!\raise.15ex\hbox {/}}
\newcommand{\slD}{\,\raise.15ex\hbox{$/$}\kern-.27em\hbox{$\!\!\!D$}}
\newcommand{\slpartial}{\raise.15ex\hbox{$/$}\kern-.57em\hbox{$\partial$}}
\newcommand{\cL}{\cal L}
\newcommand{\D}{\cal D}
\newcommand{\Dhalf}{{D\over 2}}
\def\eps{\epsilon}
\def\epshalf{{\epsilon\over 2}}
\def\lag{( -\partial^2 + V)}
\def\freeexp{{\rm e}^{-\int_0^Td\tau {1\over 4}\dot x^2}}
\def\kinb{{1\over 4}\dot x^2}
\def\kinf{{1\over 2}\psi\dot\psi}
\def\expk{{\rm exp}\biggl[\,\sum_{i<j=1}^4 G_{Bij}k_i\cdot k_j\biggr]}
\def\expp{{\rm exp}\biggl[\,\sum_{i<j=1}^4 G_{Bij}p_i\cdot p_j\biggr]}
\def\expshort{{\e}^{\half G_{Bij}k_i\cdot k_j}}
\def\expabb{{\e}^{(\cdot )}}
\def\epseps#1#2{\varepsilon_{#1}\cdot \varepsilon_{#2}}
\def\epsk#1#2{\varepsilon_{#1}\cdot k_{#2}}
\def\kk#1#2{k_{#1}\cdot k_{#2}}
\def\G#1#2{G_{B#1#2}}
\def\Gp#1#2{{\dot G_{B#1#2}}}
\def\GF#1#2{G_{F#1#2}}
\def\Dab{{(x_a-x_b)}}
\def\Dsq{{({(x_a-x_b)}^2)}}
\def\PITD{{(4\pi T)}^{-{D\over 2}}}
\def\4piTD{{(4\pi T)}^{-{D\over 2}}}
\def\4piT4{{(4\pi T)}^{-2}}
\def\TintmD{{\dps\int_{0}^{\infty}}{dT\over T}\,e^{-m^2T}
    {(4\pi T)}^{-{D\over 2}}}
\def\Tintm4{{\dps\int_{0}^{\infty}}{dT\over T}\,e^{-m^2T}
    {(4\pi T)}^{-2}}
\def\Tintm{{\dps\int_{0}^{\infty}}{dT\over T}\,e^{-m^2T}}
\def\Tint{{\dps\int_{0}^{\infty}}{dT\over T}}
\def\np{n_{+}}
\def\nm{n_{-}}
\def\Np{N_{+}}
\def\Nm{N_{-}}
\newcommand{\slG}{{{\dot G}\!\!\!\! \raise.15ex\hbox {/}}}
\newcommand{\Gd}{{\dot G}}
\newcommand{\Gund}{{\underline{\dot G}}}
\newcommand{\Gdd}{{\ddot G}}
\def\GBd12{{\dot G}_{B12}}
\def\Dx{\dps\int{\cal D}x}
\def\Dy{\dps\int{\cal D}y}
\def\Dpsi{\dps\int{\cal D}\psi}
\def\dint#1{\int\!\!\!\!\!\int\limits_{\!\!#1}}
\def\ddtau{{d\over d\tau}}
\def\ie{\hbox{$\textstyle{\int_1}$}}
\def\iz{\hbox{$\textstyle{\int_2}$}}
\def\id{\hbox{$\textstyle{\int_3}$}}
\def\ldop{\hbox{$\lbrace\mskip -4.5mu\mid$}}
\def\rdop{\hbox{$\mid\mskip -4.3mu\rbrace$}}
%
\newcommand{\1}{{\'\i}}
\newcommand{\no}{\noindent}
\def\non{\nonumber}
\def\dps{\displaystyle}
\def\sy{\scriptscriptstyle}
\def\sy{\scriptscriptstyle}

%

\newcommand{\bea}{\begin{eqnarray}}  
\newcommand{\eea}{\end{eqnarray}}  
\def\eqa{&=&}  
\def\ccr{\nonumber\\}  
  
\def\a{\alpha}
\def\b{\beta}
\def\m{\mu}
\def\n{\nu}
\def\r{\rho}
\def\s{\sigma}
\def\ep{\epsilon}

\def\cosech{\rm cosech}
\def\sech{\rm sech}
\def\coth{\rm coth}
\def\tanh{\rm tanh}

\section{Introduction}
\label{sec:intro}
\renewcommand{\theequation}{1.\arabic{equation}}
\setcounter{equation}{0}

Although processes in Einstein-Maxwell theory are presently not of direct experimental relevance, they are of great theoretical interest,
and much effort has gone into their study both at the tree-level \cite{delogi-77,atsumi-80,choi,bjerrum-03,hu-zhou-20,holstein-06} and the one-loop level \cite{bergas,bifide,frolov-96,nielsen-12,park-21}.

\begin{figure}[h]
\begin{center}
\includegraphics[scale=.7]{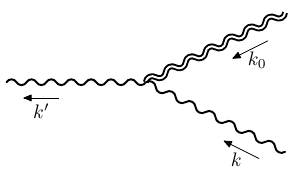}
\caption{Tree-level vertex photon-photon-graviton vertex.}
\label{fig-3vertex}
\end{center}
\end{figure}

A particularly well-studied process is photon-graviton conversion in a magnetic field.
Einstein-Maxwell theory contains a tree-level vertex involving two photons and one graviton, Fig. \ref{fig-3vertex}.

Replacing one of the photons by an electromagnetic field $F_\mn$, one can convert this vertex into the following amplitude for photon-graviton conversion
in the field (or the inverse process),
\bear
\hspace{-30pt}
\Gamma^{\rm (tree)}(k_0,\epsilon_0;k,\varepsilon;F) &=& \epsilon_{0\mn}\varepsilon_{\alpha} \Pi^{\mn,\alpha}_{\rm (tree)} (k;F)  
, \quad
 \Pi^{\mn,\alpha}_{\rm (tree)} (k;F)  = -{i \kappa\over 2} C^{\mn,\alpha}\,,
 \label{phogravtree}
\ear
where $k_0,\epsilon_0$ ($k,\varepsilon$) are the graviton (photon) momentum and polarisation, and the tensor $C^{\mn,\alpha}$ is given by
\be
C^{\mn,\alpha} =
F^{\mu\alpha}k^{\nu} +F^{\nu\alpha}k^{\mu}
-\bigl(F\cdot k\bigr)^{\mu}\delta^{\nu\alpha}
-\bigl(F\cdot k\bigr)^{\nu}\delta^{\mu\alpha} 
+ \bigl(F\cdot k\bigr)^{\alpha}\delta^{\mn} \,.
\label{defCmna}
\ee
This is well-known, but we give a derivation in appendix \ref{appA} for completeness.

%

\begin{figure}[h]
\begin{center}
\includegraphics[width=0.75\textwidth]{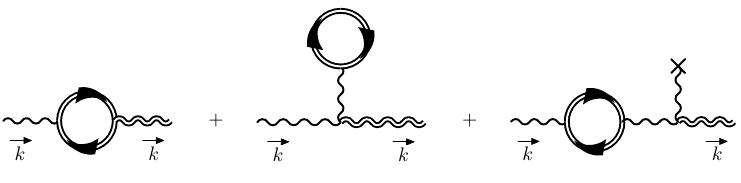}
\vskip -.7cm
$$ \hskip- .2cm \Gamma^{\rm (irr)} \hskip 4.1cm \Gamma^{\rm (tadpole)} \hskip 3.8cm   \Gamma^{\rm (ext)} $$
\caption{The three one-loop contributions to electromagnetic photon–graviton conversion. Here “irr” and “ext” denote the irreducible and ``extra'' contributions, respectively.}
\label{fig-phogravloop}

\end{center}
\end{figure}

An important application of this amplitude is the conversion of electromagnetic waves into gravitational waves (and vice-versa) in an external magnetic field,
as pointed out by Gertsenshtein as early as 1962 \cite{gertsenshtein}. 
At the tree level, these processes and the associated photon-graviton conversion have since been studied 
by many authors \cite{zelnovbook,rafsto,losotr,magueijo,chen,cilhar,defuza,eecpg}. 
However, it is only in 2005 that two of the present authors \cite{61} calculated the one-loop correction to the amplitude \eqref{phogravtree} due to the diagram shown in Fig. \ref{fig-phogravloop}
on the left, with either a scalar or spinor loop. 

\begin{figure}[h]
{\centering
\hspace{50pt}
\includegraphics[width=5.5in]{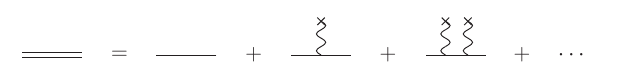}
}
\caption{Full scalar or spinor propagator in a constant field.}
\label{fig-fullprop}
\end{figure}

\noindent
Here we employ the usual double-line notation for the full propagator in the constant
external field (Fig. \ref{fig-fullprop}).

Using the then novel worldline representation of photon-graviton amplitudes
 \cite{strassler1,5,17, shaisultanov,18,41,baszir,bacozi1,bacozi2,basvan-book} 
they obtained compact parameter integral representations 
for this amplitude and for both the scalar and spinor loop cases. 
Those representations are
of the same type as the ones that one finds for the better-known 
one-loop photon vacuum polarisation in a constant field,
computed either in the worldline formalism \cite{ditsha,40} or the Feynman diagram
approach (see \cite{ditreu-book,ritus-75,ditgiebook} and refs. therein).

Although numerically these one-loop corrections for realistic field strengths are small (suppressed by a factor of $\alpha$) 
compared to the tree-level ones, in the subsequent study \cite{71} an important qualitative difference emerged: while the tree-level photon-graviton conversion amplitude 
in the field gives (contrary to the superficially similar case of photon-axion conversion \cite{maiani-86,sikivie-83,rafsto,mirizzi-06,masaki-17,chavez-14,tam-12}) equal conversion rates for both photon polarisations, at the one-loop level this rate starts becoming polarisation-dependent.
This leads to dichroism, and further analysis by Ahlers et al. \cite{ahjari}, who undertook an exhaustive search for other sources of magnetic dichroism 
showed that, although magnetic photon-graviton conversion is a very small effect under realistic conditions, remarkably it is still the leading such effect in the whole Standard Model. 
Their analysis of magnetic dichroism was also more complete in taking the right-hand diagram of Fig. \ref{fig-phogravloop} into account, which contributes to it at the same
order in $\alpha$. 

Even so, from their numbers it is clear that this effect will hardly be accessible to experiment in the near future, a conclusion that is consistent with recent detailed feasibility and sensitivity studies of light-induced gravitational effects using interferometric techniques \citep{ralf,francois}.
Nevertheless, this is of course presently the case for all process involving gravitons,
and in the long run we would still consider magnetic dichroism due to photon-graviton conversion as a leading contender in the quest for an eventual experimental demonstration of the existence of the graviton.
This is because (i) it involves only a single gravitational coupling (ii) it does not require a direct measurement of the graviton itself and (iii) dichroism can be measured with high precision using optical cavities \citep{schmitt-21,marx-13,bernhardt-20,QA-06,pvlas1,pvlas2,VMB}.

\begin{figure}[h]
\centering
\includegraphics[width=1.2in]{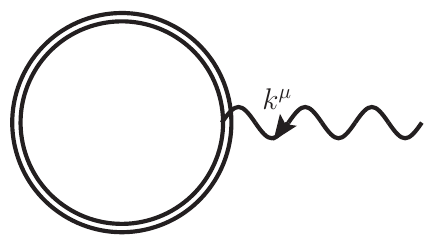}
\caption{One-photon amplitude in a constant electromagnetic field.}
\label{fig-tadpole}
\end{figure}

The main purpose of the present paper is to point out the existence of yet another diagram contributing to electromagnetic photon-graviton conversion at the same order,
 shown in the middle of Fig. \ref{fig-phogravloop}, or more precisely its non-vanishing.
This requires some explanation.
Prior to 2017 it had generally been assumed, and even stated in QED textbooks (see, e.g., \cite{ditreu-book,frgish-book}) that the one-photon tadpole diagram in a constant electromagnetic field Fig. \ref{fig-tadpole} vanishes,
and therefore also any diagram containing it.

The argument goes as follows:

\benn

\item
A constant field emits only photons with zero energy-momentum, thus there is a factor of $\delta(k)$.

\item
Because of gauge invariance, this diagram in a momentum expansion starts with the term linear in momentum.

\item
$ \delta (k) k^{\mu} =0$.

\enn

\begin{figure}[h]
\centering
\includegraphics[width=1.2in]{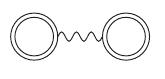}
\caption{``Handcuff'' diagram in a constant field.}
\label{fig-handcuff}
\end{figure}

Since the tadpole thus formally vanishes, it has been assumed for decades that also any diagram containing it can be
discarded, for example, the ``handcuff'' contribution Fig. \ref{fig-handcuff} to the two-loop Euler-Heisenberg Lagrangian \cite{ditreu-book,ritus-75}. 

However, in 2016 H. Gies and one of the authors \cite{giekar} noted that
such diagrams can give finite values because of the infrared divergence of the connecting photon propagator.
In dimensional regularization, the key integral is
\bear
 \int d^Dk \, \delta^D(k) \frac{k^{\mu}k^{\nu}}{k^2} = \frac{\eta^\mn}{D} \, .
 \label{intk}
\ear
Applying this integral to the handcuff diagram one finds a non-vanishing result, 
which can be expressed in the following simple way in terms of the one-loop
Euler-Heisenberg Lagrangian \cite{karbstein,karbstein-22} (see also \cite{bordag-24})
\bear
{\cal L}_{\rm spin}^{\rm 1PR}  
&=& \half \partder{{\cal L}_{\rm spin}^{(EH)}}{F^\mn} \partder{{\cal L}_{\rm spin}^{(EH)}}{F_\mn} \,,
\nonumber
\ear
(the superscript `1PR' stands for ``one-particle reducible'').

\begin{figure}[h]
\centering
\includegraphics[width=1.3in]{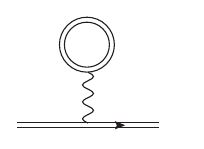}
\caption{Full scalar or spinor propagator in a constant field.}
\label{fig-proptadpole}
\end{figure}

Similarly, \cite{112,113} found that the one-loop tadpole contribution to the scalar or spinor propagator in a constant field 
(Fig.\ref{fig-proptadpole})
is also non-vanishing, and
given by
\begin{equation}
S^{\rm 1PR}(p)  = 	\frac{\partial S(p)}{\partial F_{\mu\nu}}\frac{\partial \mathcal{L}^{(EH)}}{\partial F^{\mu\nu}}  \,,
	\nonumber
\end{equation}
where $S(p)$ denotes the tree-level propagator in the field, see also \citep{ahedil-19}. 

Below we perform the similar, albeit more complicated, calculation of the middle diagram of Fig. \ref{fig-phogravloop}.
Moreover, we provide a unified calculation of all three diagrams using the worldline formalism, and show that, although the tadpole diagram contributes to the amplitude, it leaves the magnetic dichroism unchanged.
 Although our main interest is in the magnetic field and spinor loop case, we will perform all calculations in a general constant field and for both the scalar and spinor loop,
 since in the worldline formalism this essentially does not imply any extra computational effort. 
 All formulas required to specialize the field to a purely magnetic one (as well as a purely electric or
 crossed-field one) are given in appendix \ref{app-greendet}.

The organization of the paper is as follows. In section \ref{wlrep} we review the
worldline representation of the one-loop photon-graviton amplitudes with a scalar or spinor loop in vacuum,
including a short discussion of the gauge and gravitational Ward identities.
In section  \ref{sec:Inclusion of a constant electromagnetic field}
we show how to incorporate an external constant electromagnetic field. 
Section \ref{main} summarizes the worldline calculation of \cite{61,71} of the ``main'' or ``irreducible'' contributions
$\Gamma^{\rm (irr)}$ to these amplitudes, while section \ref{reducible} analyzes the contribution $\Gamma^{\rm (ext)}$. 
The central section is \ref{tadpole} where we derive the new ``tadpole'' contributions $\Gamma^{\rm (tadpole)}$. In section \ref{weak-B} we study the weak-field expansion of the tadpole contribution in a magnetic field. 
In section \ref{ward} we verify the gravitational Ward identities for these amplitudes, which also provides a non-trivial check. 
Section \ref{conc} offers our conclusions. 
Appendix \ref{appA} establishes our conventions for Einstein-Maxwell theory,
while in appendix \ref{app-greendet} we collect some useful formulas involving the
worldline Green's functions and determinants in a constant field.

Partial results of this work were presented in \cite{144}.

\section{One-loop amplitudes with one graviton and $N$ photons}
\label{wlrep}
\renewcommand{\theequation}{2.\arabic{equation}}
\setcounter{equation}{0}

We shortly summarize the worldline approach to the calculation of the 
one-loop amplitudes involving a scalar or spinor in the loop and one graviton and $N$ photons, first in vacuum
 \cite{baszir,bacozi1,bacozi2,basvan-book,87,125}. 
These amplitudes have an irreducible contribution, where all the external legs are directly attached to the loop,
and a reducible one, where the graviton is attached to one of the photons.

\subsection{Irreducible contribution}

The irreducible 
spinor/scalar loop contribution to the one-graviton $N$-photon amplitude is given by \cite{baschu-book}
\bear
\Gamma^{\rm (irr)}_{\bigl({{\rm spin}\atop{\rm scal}}\bigr)N,1}
(k_0,\epsilon_0;\ldots;k_N,\varepsilon_N)
&=&
\biggl({-2\atop 1}\biggr)
(-ie)^N \Bigl(-\frac{\kappa}{4}\Bigr)
\TintmD
\nonumber\\
&& \hskip -3cm \times
\Big \langle V^G_{\bigl({{\rm spin}\atop{\rm scal}}\bigr)} [k_0,\epsilon_0]\, 
V^\gamma_{\bigl({{\rm spin}\atop{\rm scal}}\bigr)} [k_1,\varepsilon_1] \cdots V^\gamma_{\bigl({{\rm spin}\atop{\rm scal}}\bigr)} [k_N,\varepsilon_N]  \Big \rangle \;.
\label{masterirrep}
\ear
Here $e$ and $\kappa$ are the electromagnetic and gravitational couplings, $m$ the mass and $T$ the proper-time of the loop scalar or spinor.
$k_0$ and $\epsilon_0$ denote the graviton momentum and polarization tensor, the remaining $k_i$'s and $\varepsilon_i$'s the photon ones. 
$V^G$ and $V^\gamma$ denote the graviton and photon vertex operators,
\bear
V^{\gamma}_{\rm scal}[k,\varepsilon] &=& 
  \int_0^T  d\tau\, \varepsilon\cdot \dot x(\tau) \,\e^{ik\cdot x(\tau)}\, ,\\
 V^{\gamma}_{\rm spin}[k,\varepsilon] &=& 
  \int_0^T d\tau\, \Bigl[ \varepsilon\cdot \dot x(\tau)
 -i \psi(\tau) \cdot f \cdot\psi(\tau)\Bigr] \,\e^{ik\cdot x(\tau)}
 \, ,
\label{defVspin}
\ear
\bear
V^G_{\rm scal}[k_0,\epsilon_0] &=& 
\epsilon_{0\mn}\int_0^T \!\!\! d\tau \Bigl[\dot x^{\mu}(\tau)\dot x^{\nu}(\tau)
+a^{\mu}(\tau)a^{\nu}(\tau) + b^{\mu}(\tau)c^{\nu}(\tau)
\Bigr] \e^{ik_0\cdot x(\tau)}
\, ,
\label{defVGfin}
\\
V^G_{\rm spin}[k_0,\epsilon_0] &=& 
\epsilon_{0\mn}\int_0^T \!\!\! d\tau \Bigl[
\dot x^{\mu}(\tau)\dot x^{\nu}(\tau)
+a^{\mu}(\tau)a^{\nu}(\tau) + b^{\mu}(\tau)c^{\nu}(\tau)
\nonumber\\
&&
+
\psi^\mu(\tau) \dot \psi^\nu(\tau) + \alpha^\mu(\tau) \alpha^\nu(\tau) 
+ i \dot x^{\mu}(\tau)\psi^\nu(\tau) \psi(\tau)\cdot k
\Bigr]
\e^{ik_0\cdot x(\tau)} \, ,\quad 
\label{defVgravspin}
\ear
and $f_{\mu\nu} = k_\mu \varepsilon_\nu -k_\nu \varepsilon_\mu$ the linearized photon field-strength tensor. 
The Wick contractions $\langle \cdots \rangle$ are to be performed using the set of worldline
correlators
\bear
\langle  x^{\mu}(\tau)x^{\nu}(\tau')\rangle &=& -G_B(\tau,\tau')\,, \label{wickG}\\
\langle  \psi^{\mu}(\tau)\psi^{\nu}(\tau')\rangle &=& \half G_F(\tau,\tau') \,,\label{wickGF}\\
\langle a^{\mu}(\tau)a^{\nu}(\tau')\rangle
&=&
2\delta(\tau-\tau')\delta^{\mn} \,,
\label{wicka}\\
\langle b^{\mu}(\tau)c^{\nu}(\tau') \rangle
&=&
-4 \delta(\tau-\tau')\delta^{\mn} \,,
\label{wickbc}\\
\langle \alpha^{\mu}(\tau)\alpha^{\nu}(\tau')\rangle
&=&
2\delta(\tau-\tau')\delta^{\mn}\,,
\label{wickalpha}
\label{wickrules}
\ear
with the worldline Green's functions
\bear
G_B(\tau,\tau') &=& \mid \tau - \tau'\mid - \frac{(\tau-\tau')^2}{T} \,,\label{defG}\\
G_F(\tau,\tau') &=& {\rm sgn} (\tau-\tau')\,. \label{DefGF}
\ear
\subsection{Reducible contribution}

The reducible contribution to the $N$-photon/one graviton amplitude, where the graviton is attached to one of the photons, 
usually would be constructed by sewing together the $N$-photon amplitude and the photon-photon-graviton vertex of Fig. \ref{fig-3vertex}.
For the analogous case of the scalar tree-level propagator dressed with $N$ photons and one graviton, $D_{N,1}^{({\rm red}) pp'}$, 
it was shown in \cite{125}
that the effect of this sewing can be expressed by the following replacement rule
\bear
D_{N,1}^{({\rm red}) pp'}
(k_0,\epsilon_0;\ldots;k_N,\varepsilon_N)
=
\sum_{i=1}^N 
D_{N,0}^{pp'}(k_1,\varepsilon_1;\ldots;k_i+k_0,\upsilon_i;\ldots;k_N,\varepsilon_N) \,,
\nonu
\label{17-Dred}
\ear
with an effective photon polarization vector
\bear
\upsilon_i \equiv -\kappa \frac{\lbrace \epsilon_0,f_i\rbrace\cdot (k_i+k_0)}{(k_i+k_0)^2}
=  - \kappa
\frac{\epsilon_0\cdot f_i \cdot k_0 + f_i\cdot\epsilon_0 \cdot k_i}{2k_0\cdot k_i} \,.
\label{defupsilon}
\ear 
Here we observe that the same replacement rule applies as well for the one-loop
$N$-photon/one graviton amplitudes with either a scalar or a spinor loop.
For the scalar loop this
is obvious from the fact that our derivation of that rule in \cite{125} did not make use at all 
of the boundary conditions imposed on the scalar path integral. But neither did we use the specific form of the
scalar vertex operator, and the reader can easily convince herself that changing the scalar vertex operator into
a spinor one will not make a difference here. Thus, without further ado, we can write down the reducible part
of the spinor/scalar loop contribution to the one-graviton $N$-photon amplitude as
\be
\Gamma_{\bigl({{\rm spin}\atop{\rm scal}}\bigr)N,1}^{({\rm red})}
(k_0,\epsilon_0;\ldots;k_N,\varepsilon_N)
=
\sum_{i=1}^N 
\Gamma_{\bigl({{\rm spin}\atop{\rm scal}}\bigr)N,0}
(k_1,\varepsilon_1;\ldots;k_i+k_0,\upsilon_i;\ldots;k_N,\varepsilon_N)\,,
\label{17-Gammared}
\ee
where $\upsilon_i^{\mu}$ is the same as in \eqref{defupsilon}.

\section{Inclusion of a constant electromagnetic field}
\label{sec:Inclusion of a constant electromagnetic field}
\renewcommand{\theequation}{3.\arabic{equation}}
\setcounter{equation}{0}

\subsection{General procedure}

The generalisation of the formalism to the case of the photon-graviton amplitudes in a constant electromagnetic field
is straightforward and proceeds essentially as in the purely electromagnetic case \cite{shaisultanov,18,40,41,156,lopezlopez,ahcoed}:
the master formulas \eqref{masterirrep} acquire an additional determinant factor
\bear
{\rm det}^{\half}
\biggl[
\frac{\cal Z}{{\rm tan}{\cal Z}}
\biggr]
\qquad
({\rm Spinor\,\,loop})\,, \\
{\rm det}^{\half}
\biggl[
\frac{\cal Z}{{\rm sin}{\cal Z}}
\biggr]
\qquad
({\rm Scalar\,\,loop})\,,
\label{detchange}
\ear\no
(${\cal Z}_\mn \equiv eF_\mn T$), and the worldline Green's functions $G_B,G_F$ have to be replaced by their
constant-field versions ${\cal G}_B, {\cal G}_F$,
\bear
{\cal G}_B(\tau,\tau')
&=& 
{T\over 2{\cal Z}}
\biggl(
{\e^{-i\dot G_B(\tau,\tau'){\cal Z}}-\cos{\cal Z}\over\sin{\cal Z}}
+i\dot G_B(\tau,\tau')\biggr) \;, \qquad
\nonumber\\
{\cal G}_{F}(\tau,\tau') &=&
G_F(\tau,\tau') 
{{\rm e}^{-i{\cal Z}\dot G_B (\tau,\tau')}\over {\rm cos}{\cal Z}}
\, .
\label{calGF}
\ear\no
Since these Green's functions have a non-trivial Lorentz structure, 
the Wick-contraction rules \eqref{wickG}, \eqref{wickGF} generalise in the form
\begin{eqnarray}
\langle x^{\mu}(\tau)x^{\nu}(\tau')\rangle_F
&=&
-{\cal G}_B^{\mu\nu}(\tau,\tau') \, ,
\label{wickcalGB}\\
\langle\psi^{\mu}(\tau)\psi^{\nu}(\tau')\rangle_F
&=&
\frac{1}{2}{\cal G}_F^{\mu\nu}(\tau,\tau')\,,
\label{wickcalGF} 
\end{eqnarray}
with the Wick-contraction rules for the ghosts unchanged. 

Thus the vacuum master formula \eqref{masterirrep} directly generalises to the constant-field background as
\bear
\Gamma^{\rm (irr)}_{\bigl({{\rm spin}\atop{\rm scal}}\bigr)N,1}
(k_0,\epsilon_0;\ldots;k_N,\varepsilon_N;F)
&=&
\biggl({-2\atop 1}\biggr)
(-ie)^N \Bigl(-\frac{\kappa}{4}\Bigr)
\TintmD
{\rm det}^{\half}
\biggl[
\biggl(
{{\cal Z}/{\rm tan}{\cal Z}\atop {\cal Z}/{\rm sin}{\cal Z}}
\biggr)
\biggr]
\nonumber\\
&& \hskip -3cm \times
\Big \langle V^G_{\bigl({{\rm spin}\atop{\rm scal}}\bigr)} [k_0,\epsilon_0]\, 
V^\gamma_{\bigl({{\rm spin}\atop{\rm scal}}\bigr)} [k_1,\varepsilon_1] \cdots V^\gamma_{\bigl({{\rm spin}\atop{\rm scal}}\bigr)} [k_N,\varepsilon_N]  \Big \rangle_F \;.
\label{masterirrepF}
\ear

The rule \eqref{17-Gammared} for the construction of the reducible contributions, where the graviton
gets attached to one of the photons, is also not affected by the presence of the external field.

\begin{figure}[h]
\begin{center}
 \includegraphics[width=0.25\textwidth]{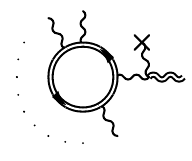}
\caption{{Field-induced type of reducible contributions.}}
\label{fig-secondreducible}
\end{center}
\end{figure}

At the one-graviton level, the only really new aspect arising from the presence of the external field is that there is now a 
second type of reducible contributions, since the graviton can now be attached also to one of the photons emitted by the external field,
see  Fig. \ref{fig-secondreducible}. 

It is easy to see that the contribution of this diagram, to be called 
$\Gamma_{N,1}^{({\rm ext})}$, 
can be obtained from the purely photonic amplitude with
$N+1$ photons in the field by the following modification of \eqref{17-Gammared},
\be
\Gamma_{\bigl({{\rm spin}\atop{\rm scal}}\bigr)N,1}^{({\rm ext})}
(k_0,\epsilon_0;\ldots;k_N,\varepsilon_N;F)
=
\Gamma_{\bigl({{\rm spin}\atop{\rm scal}}\bigr)N+1,0}
(k_0,\upsilon_F;k_1,\varepsilon_1;\ldots;\ldots;k_N,\varepsilon_N;F)\,,
\label{17-Gammaext}
\ee
where now
\be
\upsilon_F \equiv i \kappa \frac{\lbrace \epsilon_0,F\rbrace\cdot k_0}{k_0^2} \;.
\label{defupsilonF}
\ee

\subsection{The photon tadpole}
\label{subsec:The photon tadpole}

For starters, let us work out \eqref{masterirrepF} for the pure QED case with one and two photons. 
For the tadpole (Fig. \ref{fig-tadpole}) with a scalar loop, this yields
\begin{eqnarray}
\Gamma_{\rm scal}
(k,\varepsilon;F)
&=&
-e
{(2\pi )}^D\delta^D (k)
\int_{0}^{\infty}}{dT\over T}
{(4\pi T)^{-{D\over 2}}
e^{-m^2T}
{\rm det}^{{1\over 2}} \biggl[\frac{\cZ}{\sin\cZ}\biggr]
\nonumber\\ 
&& \times
\int_0^T  d\tau \,
\varepsilon\cdot\dot{\cal G}_{B}(\tau,\tau)\cdot k
\,{\rm exp}\Bigl[\half k\cdot {\cal G}_{B}(\tau,\tau)\cdot  k\Bigr] \;.
\label{1photonscalT}
\end{eqnarray}
Using the coincidence limits \eqref{app-gd-coincalGB}, \eqref{app-gd-coindcalGB} this can be written more
explicitly as
\begin{eqnarray}
\Gamma_{\rm scal}
(k,\varepsilon;F)
&=&
-ie
{(2\pi )}^D\delta^D (k)
\int_{0}^{\infty}dT
{(4\pi T)}^{-{D\over 2}}
e^{-m^2T}
{\rm det}^{\frac{1}{2}} \biggl[ \frac{\cZ}{\sin \cZ}\biggr]
\nonumber\\ 
&& \times
\varepsilon\cdot
\Bigl({\cot}{\cZ} - \frac{1}{\cZ}\Bigr)
\cdot k
\,{\rm exp}\biggl[\frac{T}{4} 
k\cdot 
\Bigl(\frac{{\cot}\cZ}{\cZ} - \frac{1}{{\cZ}^2}\Bigr)
\cdot  k\biggr] \;.
\label{1photonscalTfin}
\end{eqnarray}
To get the corresponding quantity for spinor QED, we have to include the spin part of the photon vertex operator,
change the determinant factor, and to supply the usual
global factor of $-2$. This yields 
\begin{eqnarray}
\Gamma_{\rm spin}
(k,\varepsilon;F)
&=&
2ie
{(2\pi )}^D\delta^D (k)
\int_{0}^{\infty}}dT
{(4\pi T)^{-{D\over 2}}
e^{-m^2T}
{\rm det}^{{1\over 2}} \biggl[\frac{\cZ}{\tan\cZ}\biggr]
\nonumber\\ 
&& \hskip -1.5cm
\times
\varepsilon\cdot
\Bigl({\cot}{\cZ} + {\tan}{\cZ}  - \frac{1}{\cZ}\Bigr)
\cdot k
\,{\rm exp}\biggl[\frac{T}{4} 
k\cdot 
\Bigl(\frac{{\cot}\cZ}{\cZ}- \frac{1}{{\cZ}^2}\Bigr)
\cdot  k\biggr] \;.
\label{1photonspinTfin}
\end{eqnarray}
Note that, as mentioned in the introduction, the expressions \eqref{1photonscalTfin}, \eqref{1photonspinTfin} formally vanish
on account of the factor $\delta^D(k)\, k^{\mu}$. 

\subsection{The vacuum-polarization tensor in a constant field}
\label{subsec:The vacuum polarization tensor in a constant field}

For $N=2$ in the scalar case we find the integrand
\bear
\Big \langle V^\gamma_{\rm scal}[k_1,\varepsilon_1] V^\gamma_{\rm scal}[k_2,\varepsilon_2] \Big \rangle_F 
&=&
\Bigl\lbrack
\varepsilon_1\cdot\ddot {\cal G}_{B12}\cdot\varepsilon_2
-
\varepsilon_1\cdot{ {\dot{\cal G}}}_{B1i}\cdot k_i
\,\varepsilon_2\cdot{ {\dot{\cal G}}}_{B2j}\cdot k_j
\Bigr\rbrack
\e^{\half k_i\cdot {\cal G}_{Bij}\cdot k_j}\,,
\nonumber\\
\label{P2withF}
\ear
where summation over $i,j = 1,2$ is understood.
%
%
We use momentum conservation to write $k=k_1= -k_2$.
Removing the second derivative in the first term by an IBP in $\tau_1$, the integrand becomes
\be
\Bigl\lbrack
\varepsilon_1\cdot{{\dot {\cal G}}}_{B12}\cdot\varepsilon_2
\,k\cdot {\dot{\cal G}}_{B12}\cdot k
+
\varepsilon_1\cdot
\Bigl(\dot{\cal G}_{B12}-\dot{\cal G}_B\Bigr)
\cdot k
\,\varepsilon_2\cdot
\Bigl(
\dot{\cal G}_{B21} - \dot{\cal G}_B
\Bigr)
\cdot k
\Bigr\rbrack
\e^{-k\cdot ({\cal G}_{B12}- {\cal G}_B)\cdot k} \,,
\label{P2withFpint}
\ee
where we have introduced the further notation of 
indicating the constant coincidence limit of a Green's function by the omission of its
argument, $\dot{\cal G}_{B} \equiv \dot{\cal G}_{Bii}$ etc. 
The content of the brackets then turns into $\varepsilon_{1\mu} I_{\rm scal}^{\mu\nu}\varepsilon_{2\nu}$,
\bear
I^{\mu\nu}_{\rm scal} &=&
\dot{\cal G}^{\mu\nu}_{B12}k\cdot\dot{\cal G}_{B12}\cdot k
+
\Bigl(\dot{\cal G}_{B12}
-\dot{\cal G}_{B}\Bigr)
^{\mu\lambda}
  \Bigl(\dot{\cal G}_{B21}
-\dot{\cal G}_{B}
\Bigr)^{\nu\kappa}
k^{\kappa}k^{\lambda}
\, .
\label{Iscal}
\ear\no
Next we would like to use the fact that the integrand
contains terms which integrate to zero
due to antisymmetry
under the exchange
$\tau_1\leftrightarrow\tau_2$. 
Thus we decompose the constant-field worldline Green's function as
\bear
{\cal G}_{B12} = {\cal S}_{B12} + {\cal A}_{B12}\,,
\label{evenoddcalGB}
\ear
where ${\cal S}_{Bij}$ denotes the even (${\cal S}_{Bii}\equiv {\cal S}_B$) and ${\cal A}_{Bij}$ the odd part (${\cal A}_{Bii}\equiv {\cal A}_B$) as a power series in $F$.
Only the Lorentz even part
of ${\cal G}_{Bij}$ contributes in the exponent,
\bear
k\cdot ({\cal G}_{B12}-{\cal G}_B) \cdot k
&=&
k\cdot \bigl( {\cal S}_{B12}-{\cal S}_{B}\bigr)\cdot k
\equiv
Tk\cdot\Phi_{12}\cdot k
\label{defPhi} \;.
\ear\no
$I^{\mu\nu}_{\rm scal}$ turns, after decomposing all factors of $\dot {\cal G}_{Bij}$ as above, and deleting all $\tau$ - odd terms, into 
\bear
\tilde I^{\mu\nu}_{\rm scal}
&\equiv&
\biggl\lbrace
{\dot{\cal S}}^{\mu\nu}_{B12}{\dot{\cal S}}^{\kappa\lambda}_{B12}
- {\dot{\cal S}}^{\mu\lambda}_{B12}{\dot{\cal S}}^{\nu\kappa}_{B12}
+
\Bigl(
{\dot{\cal A}}_{B12}-{\dot{\cal A}}_{B}
\Bigr)^{\mu\lambda}
\Bigl(
{\dot{\cal A}}_{B12}-{\dot{\cal A}}_{B}
\Bigr)^{\nu\kappa}
\biggr\rbrace
k^{\kappa}k^{\lambda}
\, .
\nonumber\\
\label{intfinal}
\ear\no
In this way we obtain the following integral representations 
for the dimensionally regularized constant-field vacuum polarization tensor in scalar QED:
\bear
\Pi^{\mu\nu}_{\rm scal}(k;F)
&=&
-{e^2\over {(4\pi)}^{D\over 2}}
{\dps\int_{0}^{\infty}}{dT\over T}
{T}^{2-{D\over 2}}
e^{-m^2T}
\detsin 
\int_0^1 du_1
\, \tilde I^{\mu\nu}_{\rm scal}
\,\e^{-Tk\cdot\Phi_{12}\cdot k}
\, .
\nonumber\\
\label{vpscalreg}\ear\no
As usual we have rescaled to the unit circle and set $u_2 =0$. 

For spinor QED, the formula corresponding to \eqref{Iscal} is \cite{40}
\bear
I^{\mu\nu}_{\rm spin} &=&
\dot{\cal G}^{\mu\nu}_{B12}k\cdot\dot{\cal G}_{B12}\cdot k
-{\cal G}^{\mu\nu}_{F12}k\cdot{\cal G}_{F12}\cdot k
\label{substint}
\\&&\hspace{-20pt}
- \biggl[
\Bigl(\dot{\cal G}_{B}
-{\cal G}_{F}
-\dot{\cal G}_{B12}\Bigr)
^{\mu\lambda}
  \Bigl(\dot{\cal G}_{B21}
-\dot{\cal G}_{B}
+{\cal G}_{F}
\Bigr)^{\nu\kappa}
+
{\cal G}^{\mu\lambda}_{F12}
{\cal G}^{\nu\kappa}_{F21}
\biggr]
k^{\kappa}k^{\lambda} 
\nonumber
\ear\no
and \eqref{intfinal} generalizes to 
\bear
\tilde I^{\mu\nu}_{\rm spin}
&\equiv&
\biggl\lbrace
\Bigl(
{\dot{\cal S}}^{\mu\nu}_{B12}{\dot{\cal S}}^{\kappa\lambda}_{B12}
- {\dot{\cal S}}^{\mu\lambda}_{B12}{\dot{\cal S}}^{\nu\kappa}_{B12}
\Bigr)
-
\Bigl(
{{\cal S}}^{\mu\nu}_{F12}{{\cal S}}^{\kappa\lambda}_{F12}
- {{\cal S}}^{\mu\lambda}_{F12}{{\cal S}}^{\nu\kappa}_{F12}
\Bigr)
\nonumber\\
&&
+
\Bigl(
{\dot{\cal A}}_{B12}-{\dot{\cal A}}_{B}+{{\cal A}}_{F}
\Bigr)^{\mu\lambda}
\Bigl(
{\dot{\cal A}}_{B12}-{\dot{\cal A}}_{B}+{{\cal A}}_{F}
\Bigr)^{\nu\kappa}
\nonumber\\
&&
-{\cal A}^{\mu\lambda}_{F12}
{\cal A}^{\nu\kappa}_{F12}
\biggr\rbrace
k^{\kappa}k^{\lambda} 
\, .
\label{intfinalspin}
\ear\no
Taking the global factor of $-2$ and the change of the field-dependent determinant factor \eqref{detchange} into account, 
\eqref{vpscalreg} generalizes to 
\bear
\Pi^{\mu\nu}_{\rm spin}(k;F)
&=&
2{e^2\over {(4\pi)}^{D\over 2}}
{\dps\int_{0}^{\infty}}{dT\over T}
{T}^{2-{D\over 2}}
e^{-m^2T}
\dettan
\int_0^1 du_1
\, \tilde I^{\mu\nu}_{\rm spin}
\,\e^{-Tk\cdot\Phi_{12}\cdot k}
\, .
\nonumber\\
\label{vpspinreg}
\ear\no
As is well-known, the constant field vacuum polarization tensors contain
the UV divergences of the ordinary vacuum polarization tensors,
but no further ones, so that on-shell renormalization can be performed by
subtraction of the integrand at zero field strength and momentum. 
In this way we find for the renormalized vacuum polarization tensors 
(indicating the renormalisation by a `bar') \cite{40}
\bear
\bar\Pi^{\mu\nu}_{\rm scal}(k;F)
&=&
-\frac{e^2}{16\pi^2}
{\dps\int_{0}^{\infty}}{dT\over T}
e^{-m^2T}
\int_0^1 du_1
\biggl\lbrace
\detsin
\, \tilde I^{\mu\nu}_{\rm scal}
\,\e^{-Tk\cdot\Phi_{12}\cdot k}
\nonumber\\&&\hspace{100pt}
 - \bigl(\delta^{\mu\nu}k^2 -k^{\mu}k^{\nu}\bigr)(1-2u_1)^2
\biggr\rbrace
\, ,
\label{vpscalren}\\
\bar\Pi^{\mu\nu}_{\rm spin}(k;F)
&=&
\frac{e^2}{8\pi^2}
{\dps\int_{0}^{\infty}}{dT\over T}
e^{-m^2T}
\int_0^1 du_1
\biggl\lbrace
\dettan
\, \tilde I^{\mu\nu}_{\rm spin}
\,\e^{-Tk\cdot\Phi_{12}\cdot k}
\nonumber\\&&
\hspace{100pt}
 +4 \bigl(\delta^{\mu\nu}k^2 -k^{\mu}k^{\nu}\bigr)u_1(1-u_1)
\biggr\rbrace \;.
\label{vpspinren}
\ear\no
The remaining $u_1$ - integral can be brought
into a more standard form
by a transformation of variables
$v = \dot G_{B12} = 1-2u_1$.

The matrix decomposition formulas of appendix \ref{app-greendet} can be used to write these integral representations more explicitly
(for the generic case of parallel electric and magnetic fields this has been done in \cite{40}). 

\section{Irreducible contribution to electromagnetic photon-graviton conversion}
\label{main}
\renewcommand{\theequation}{4.\arabic{equation}}
\setcounter{equation}{0}

We now turn to applying this formalism to the electromagnetic photon-graviton conversion amplitude,
starting with the irreducible contribution $\Gamma_{\rm spin}^{\rm (irr)}$.
Using the machinery developed above, its worldline representation can be written as
\bear
\Gamma_{\rm spin}^{\rm (irr)}(k_0,\epsilon_0;k,\varepsilon;F) &=& 
-2
(-ie)(-{\kappa\over 4}) 
\TintmD 
{\rm det}^{-{1\over 2}}
\biggl[{{\rm tan}({\cal Z})\over {\cal Z}}\biggr]
\nonumber\\&&\hspace{50pt}
\times
\Bigl\langle
V^G_{\rm spin}[k_0,\epsilon_0] V^\gamma_{\rm spin}[k,\varepsilon]
\Bigr\rangle_F \,.
\label{hAwlwickf} 
\ear
Performing the Wick contractions according to the rules for the constant-field case, eqs.
\eqref{wickcalGB} and \eqref{wickcalGF}, stripping off $\epsilon_0,\varepsilon$ as in \eqref{phogravtree}, 
and eliminating $k_0$ through $k_0=-k$, the result can be written as
\bear
\Pi^{\mu\nu ,\alpha}_{\rm spin}(k;F) &=&
-{e\kappa  \over 2 (4\pi)^{D\over 2}}
\Tintm T^{-{D\over 2}}
\,
{\rm det}^{-{1\over 2}}
\Bigl[{{\rm tan}{\cal Z}\over {\cal Z}}\Bigr]
\nonumber\\&&\times
\int_0^Td\tau_1\int_0^Td\tau_2
\,\e^{-k\cdot \bar{\cal G}_{B12}\cdot k}
\bigl(J^{\mu\nu ,\alpha}_{\rm scal} +J_{\psi}^{(\mn),\alpha}\bigr)\,,
\label{hAwlwickresultspin} 
\ear
where in $J^{\mu\nu ,\alpha}_{\rm scal}$ we have collected all terms that come from the purely orbital part of the vertex operators, and
$J_{\psi}^{(\mn),\alpha}\equiv \half \bigl(J_{\psi}^{(\mn),\alpha} + J_{\psi}^{(\nu\mu ),\alpha}\bigr)$ 
contains all the terms involving the spin of the loop fermion. Here $\bar{\mathcal G}_{B12} \equiv \mathcal G_{B12}-\mathcal G_B$ is the modified Green’s function obtained by subtracting its coincidence limit.
Using the symmetry properties  \eqref{app-gd-symmcalGB} and \eqref{app-gd-symmcalGF}, the result can be written as
\bear
J^{\mu\nu ,\alpha}_{\rm scal}
&=&
\Bigl({\ddot {\cal G}}_{B11}^{\mn}-2\delta_{11}\delta^{\mn}\Bigr)
\Bigl(k\cdot \,{\overline {\dot {\cal G}}}_{B12}\Bigr)^{\alpha}
+\Bigl[ {\ddot {\cal G}}_{B12}^{\mu\alpha}
\Bigl(\,{\overline {\dot {\cal G}}}_{B12}\cdot k\Bigr)^{\nu}
+ \exmn \Bigr] \nonumber\\&&
- \Bigl(\,{\overline {\dot {\cal G}}}_{B12}\cdot k\Bigr)^{\mu}
\Bigl(\,{\overline {\dot {\cal G}}}_{B12}\cdot k\Bigr)^{\nu}
\Bigl(k\cdot \,{\overline {\dot {\cal G}}}_{B12}\Bigr)^{\alpha} \;,
\label{Imunualpha}
\ear
\bear
J^{\mu\nu ,\alpha}_{\psi}
&=&
\Bigl[{\dot {\cal G}}_{F11}^{\mn}-2\delta_{11}\delta^{\mn}
+
\Bigl(\,{ { {\cal G}}}_{F11}\cdot k\Bigr)^{\nu}
\Bigl(\,{\overline {\dot {\cal G}}}_{B12}\cdot k\Bigr)^{\mu}
\Bigr]
\Bigl(\,{ { {\cal G}}}_{F22}\cdot k\Bigr)^{\alpha}
\nonumber\\&&
-
\Bigl(\,{ { {\cal G}}}_{F12}\cdot k\Bigr)^{\mu}
{ {\dot {\cal G}}}_{F12}^{\nu \alpha}
+
{ { {\cal G}}}_{F12}^{\mu\alpha}
\Bigl(\,{ {\dot {\cal G}}}_{F12}\cdot k\Bigr)^{\nu}
\nonumber\\&&
+\Bigl[-{{\cal G}}_{F12}^{\nu\alpha}
\Bigl(\,{ {k\cdot {\cal G}}}_{F12}\cdot k\Bigr)
+
\Bigl(\,{ { {\cal G}}}_{F12}\cdot k\Bigr)^{\nu}
\Bigl(\,{ {k\cdot {\cal G}}}_{F12}\Bigr)^\alpha \Bigr]
\Bigl(\,{\overline {\dot {\cal G}}}_{B12}\cdot k\Bigr)^{\mu}
\nonumber\\&&
-
\Bigl({\dot {\cal G}}_{F11}^{\mn}-2\delta_{11}\delta^{\mn}\Bigr)
\Bigl(\, k\cdot {\overline {\dot {\cal G}}}_{B12}\Bigr)^{\alpha}
\nonumber\\&&
- \Bigl[
\ddot {\cal G}_{B11}^{\mu\nu}
-2\delta_{11}\delta^{\mn} -
\Bigl(\,{\overline {\dot {\cal G}}}_{B12}\cdot k\Bigr)^{\mu}
\Bigl(\,{\overline {\dot {\cal G}}}_{B12}\cdot k\Bigr)^{\nu}
\Bigr]
\Bigl(\,{ { {\cal G}}}_{F22}\cdot k\Bigr)^{\alpha} 
\nonumber\\&&
+\Bigl [ 
{\ddot {\cal G}}_{B12}^{\mu\alpha} 
- \Bigl(\,{\overline {\dot {\cal G}}}_{B12}\cdot k\Bigr)^{\mu}
\Bigl(\, k\cdot {\overline {\dot {\cal G}}}_{B12}\Bigr)^{\alpha}
\Bigr ]
\Bigl(\,{ { {\cal G}}}_{F11}\cdot k\Bigr)^{\nu} \, .
\label{Iextra}
\ear
It will be useful to add to $J_{\rm scal}^{\mu\nu ,\alpha}\,\e^{-k\cdot \overline {\cal G}_{B12}\cdot k}$ 
the total derivative term
\bear
-\half {\partial\over \partial\tau_1}
\Bigl[
\,{\overline {\dot {\cal G}}}_{B12}^{\mu\alpha}
\Bigl(\,{\overline {\dot {\cal G}}}_{B12}\cdot k\Bigr)^{\nu}
\,\e^{-k\cdot \overline {\cal G}_{B12}\cdot k}
+ \exmn
\Bigr].
\label{addtotder}
\ear
We can then replace $J^{\mu\nu ,\alpha}_{\rm scal} +J_{\psi}^{(\mn),\alpha}$ by 
\bear
J_{\rm spin}^{\mn ,\alpha}
&=&
J_{{\rm spin} ,1}^{\mn ,\alpha}
+J_{{\rm spin} ,2}^{(\mn) ,\alpha}
+J_{{\rm spin} ,3}^{(\mn) ,\alpha}
\label{decomposeIspin}
\ear
where
\bear
J_{{\rm spin} ,1}^{\mn ,\alpha}
&=& -
\Bigl({\ddot {\cal G}}_{B11}^{\mn}-{\dot{\cal G}}_{F11}^{\mn}\Bigr)
\Bigl[
\bigl({\overline {\dot {\cal G}}}_{B21}+
{\cal G}_{F22}\bigr)\cdot k\Bigr]^{\alpha} \, ,
\nonumber\\
J_{{\rm spin} ,2}^{\mn ,\alpha}
&=&
-{\overline{\dot {\cal G}}}_{B12}^{\mu\alpha}
\Bigl( \ddot {\cal G}_{B12}\cdot k\Bigr)^{\nu}
+{\cal G}_{F12}^{\mu\alpha}
\Bigl(\dot {\cal G}_{F12}\cdot k\Bigr)^{\nu}
+{\ddot {\cal G}}_{B12}^{\nu\alpha}
\Bigl[\bigl({\overline {\dot {\cal G}}}_{B12}+{\cal G}_{F11}\bigr)\cdot k\Bigr]^{\mu} 
\nonumber\\&&
-{\dot {\cal G}}_{F12}^{\nu\alpha}
\Bigl({\cal G}_{F12}\cdot k\Bigr)^{\mu} \, ,
\nonumber\\
J_{{\rm spin} ,3}^{\mn ,\alpha}
&=&
\,\Bigl({\overline{\dot {\cal G}}}_{B12}\cdot k\Bigr)^{\mu}
\Bigl\lbrace
\Bigl[\bigl({\overline{\dot{\cal G}}}_{B12}+{\cal G}_{F11}\bigr)\cdot k\Bigr]^{\nu}
\Bigl[\bigl({\overline{\dot{\cal G}}}_{B21}+{\cal G}_{F22}\bigr)\cdot k\Bigr]^{\alpha}
\nonumber\\&&
-\Bigl({\cal G}_{F12}k\Bigr)^{\nu}
\Bigl({\cal G}_{F21}k\Bigr)^{\alpha}
+{\overline{\dot{\cal G}}}_{B12}^{\nu\alpha}k\cdot {\dot{\cal G}}_{B12}\cdot k 
-{\cal G}_{F12}^{\nu\alpha}k\cdot {\cal G}_{F12}\cdot k
\Bigr\rbrace.
\label{Iispin}
\ear
Before proceeding further, let us use this integral representation
to analyse the general structure of this amplitude. Although
our calculation is nonperturbative in the external field,
we can use the series expansions of ${\cal G}_B,\dot{\cal G}_B,\ddot{\cal G}_B$ and ${\cal G}_F,\dot{\cal G}_F$ in the field, 
whose first few terms are given in \eqref{app-gd-GBexpand} - \eqref{app-gd-ddGBexpand} and \eqref{app-gd-GFexpand}, \eqref{app-gd-dGFexpand},
to compute the amplitude involving a given number of interactions with the field.
It is then immediately seen that this amplitude is nonzero only
if this number of interactions is odd, since otherwise the
$\tau_{1,2}$ integrations vanish by antisymmetry. 

The leading term in this expansion in the background field is UV divergent.
In dimensional regularisation, this divergence is \cite{61}
\bear
\Pi_{\rm spin, div}^{\mn,\alpha}(k;F)
&=& - \frac{4}{3}
{ie^2\kappa\over(4\pi)^2}{1\over D-4}C^{\mn,\alpha}(k)\,,
\label{Adiv}
\ear
where $C^{\mn,\alpha}$ is the same tensor that appeared in the tree-level amplitude, eq. (\ref{defCmna}).
This was to be expected since the amplitude must be multiplicatively renormalizable
(although quantum gravity is not a renormalizable theory, 
amplitudes in Einstein-Maxwell theory involving gravitational fields and gravitons only externally are still renormalizable
by power counting). 
Renormalization is performed by subtracting the amplitude
at zero field strength and zero momentum, leading to the following form of the renormalized
amplitude $\bar \Pi$:
\bear
{\bar \Pi}^{\mu\nu ,\alpha}_{\rm spin}(k;F) &=&
-{e\kappa \over  32 \pi^2}
\int_0^{\infty}{dT\over T^3}
\,\e^{-m^2T}\
\\&&
\hspace{-6mm}
\times \biggl\lbrace
{\rm det}^{-{1\over 2}}
\Bigl[{{\rm tan}({\cal Z})\over {\cal Z}}\Bigr]
\int_0^T \!\!\! d\tau_1\int_0^T \!\!\! d\tau_2
\,\e^{-k\cdot \overline {\cal G}_{B12}\cdot k}
J^{\mu\nu ,\alpha}_{\rm spin}
+{4\over 3}iT^2 e C^{\mn,\alpha}
\biggr\rbrace . 
\nonumber
\ear
Once more this can still be simplified using the decompositions 
%
$
{\cal G}_{B,F} = {\cal S}_{B,F} + {\cal A}_{B,F}
$
where ${\cal S}_{B,F}^{\mu\nu}$ contains the even powers of $F^{\mu\nu}$
in the power series representation of ${\cal G}_{B,F}^{\mu\nu}$, and
${\cal A}_{B,F}^{\mu\nu}$ the odd ones. After this replacement
all terms in the integrand are either symmetric or antisymmetric
under the exchange $\tau_1 \leftrightarrow \tau_2$, and the
antisymmetric ones can be deleted since their $\tau_{1,2}$ - integrals
vanish. 
Further, as usual, we rescale $\tau_i = Tu_i, i=1,2$, and set $u_2=0$. This leads to
\bear
{\bar \Pi}^{\mu\nu ,\alpha}_{\rm spin}(k;F) &=&
-{e\kappa\over 32 \pi^2}
\int_0^{\infty}{dT\over T}
\,\e^{-m^2T}
\Biggl\lbrace
{\rm det}^{-{1\over 2}}
\Bigl[{{\rm tan}({\cal Z})\over {\cal Z}}\Bigr]
\int_{0}^1 \!\!\!
du_1
\nonumber\\&&\hspace{40pt}\times
\,\e^{-k\cdot {\overline{\cal S}_{B12}}\cdot k}
\sum_{m=1}^3
{\tilde J}^{(\mu\nu),\alpha}_{\rm spin,m}
+\frac{4}{3}ieC^{\mu\nu ,\alpha}
\Biggr\rbrace\,,
\label{Aspinfin} 
\ear
where
\bear
\tilde J^{\mu\nu,\alpha}_{{\rm spin}, 1} &=&
\Bigl({\ddot {\cal S}}_{B11}^{\mn}-{\dot {\cal S}}_{F11}^{\mn})
\Bigl(k\cdot \,\bigl({\overline {\dot {\cal A}}}_{B12}+{\cal A}_{F22}\bigr)\Bigr)^{\alpha} \, ,
\nonumber\\
\tilde J^{\mu\nu,\alpha}_{{\rm spin}, 2} &=&
-{\dot {\cal S}}_{B12}^{\mu\alpha}
\Bigl( \ddot {\cal A}_{B12}\cdot k\Bigr)^{\nu}
+
{\cal S}_{F12}^{\mu\alpha}
\Bigl( \dot {\cal A}_{F12}\cdot k\Bigr)^{\nu}
+{\ddot {\cal A}}_{B12}^{\nu\alpha}
\Bigl(\,{\dot {\cal S}}_{B12}\cdot k\Bigr)^{\mu}
-
{\dot {\cal A}}_{F12}^{\nu\alpha}
\Bigl( {\cal S}_{F12}\cdot k\Bigr)^{\mu}
\nonumber\\&&
-\,{\overline{\dot {\cal A}}}_{B12}^{\mu\alpha}
\Bigl( \ddot {\cal S}_{B12}\cdot k\Bigr)^{\nu}
+
{\cal A}_{F12}^{\mu\alpha}
\Bigl( \dot {\cal S}_{F12}\cdot k\Bigr)^{\nu}
\nonumber\\&&
+{\ddot {\cal S}}_{B12}^{\nu\alpha}
\Bigl(\,\bigl({\overline {\dot {\cal A}}}_{B12}+{\cal A}_{F11}\bigr)\cdot k\Bigr)^{\mu}
-
{\dot {\cal S}}_{F12}^{\nu\alpha}
\Bigl({\cal A}_{F12}\cdot k\Bigr)^{\mu} \, ,
\nonumber\\
\tilde J^{\mu\nu,\alpha}_{{\rm spin}, 3} &=&
-
\Bigl(\,{\dot {\cal S}}_{B12}\cdot k\Bigr)^{\mu}
\Bigl\lbrack
\Bigl(\,{\dot {\cal S}}_{B12}\cdot k\Bigr)^{\nu}
\Bigl(k\cdot \, \bigl({\overline {\dot {\cal A}}}_{B12}+{\cal A}_{F11}\bigr)\Bigr)^{\alpha}
-
\Bigl(\,{\cal S}_{F12}\cdot k\Bigr)^{\nu}
\Bigl(k\cdot  {\cal A}_{F12} \Bigr)^{\alpha}
\nonumber\\&&
+
\Bigl(\, \bigl({\overline  {\dot {\cal A}}}_{B12}+{\cal A}_{F11}\bigr)\cdot k\Bigr)^{\nu}
\Bigl(k\cdot \,{\dot {\cal S}}_{B12}\Bigr)^{\alpha}
-
\Bigl({\cal A}_{F12}\cdot k\Bigr)^{\nu}
\Bigl(k\cdot \,{\cal S}_{F12}\Bigr)^{\alpha}
\nonumber\\&&\hspace{65pt}
-
\,{\overline{\dot {\cal A}}}_{B12}^{\nu\alpha}
k\cdot {\dot {\cal S}}_{B12}\cdot k
+
{\cal A}_{F12}^{\nu\alpha}
k\cdot {\cal S}_{F12}\cdot k
\Bigr\rbrack
\nonumber\\&&
-
\Bigl(\,{\overline {\dot {\cal A}}}_{B12}\cdot k\Bigr)^{\mu}
\Bigl\lbrack
\Bigl(\,{\dot {\cal S}}_{B12}\cdot k\Bigr)^{\nu}
\Bigl(k\cdot \,          {\dot {\cal S}}_{B12}\Bigr)^{\alpha}
-
\Bigl({\cal S}_{F12}\cdot k\Bigr)^{\nu}
\Bigl(k\cdot \, {\cal S}_{F12}\Bigr)^{\alpha}
\nonumber\\&&
+
\Bigl(\, \bigl({\overline  {\dot {\cal A}}}_{B12}+{\cal A}_{F11}\bigr)\cdot k\Bigr)^{\nu}
\Bigl(k\cdot \, \bigl({\overline {\dot {\cal A}}}_{B12}+{\cal A}_{F11}\bigr) \Bigr)^{\alpha}
-
\Bigl({\cal A}_{F12}\cdot k\Bigr)^{\nu}
\Bigl(k\cdot \, {\cal A}_{F12}\Bigr)^{\alpha}
\nonumber\\&&\hspace{65pt}
-
\,{\dot {\cal S}}_{B12}^{\nu\alpha}
k\cdot {\dot {\cal S}}_{B12}\cdot k
+
{\cal S}_{F12}^{\nu\alpha}
k\cdot {\cal S}_{F12}\cdot k
\Bigr\rbrack \, .
\label{Jtildespin} 
\ear
It is then straightforward to write the integrands more explicitly using the decomposition formulas of appendix \ref{app-greendet}.

The corresponding result for the scalar loop is obtained by omitting all terms involving the fermionic
worldline Green's function, changing ${\rm det}^{-{1\over 2}}\Bigl[{{\rm tan}({\cal Z})\over {\cal Z}}\Bigr]$
to
${\rm det}^{-{1\over 2}}\Bigl[{{\rm sin}({\cal Z})\over {\cal Z}}\Bigr]$,
and multiplying by a global factor of $-\half$. 

\section{Tadpole contribution to electromagnetic photon-graviton conversion}
\label{tadpole}
\renewcommand{\theequation}{5.\arabic{equation}}
\setcounter{equation}{0}

The tadpole contribution (middle diagram in Fig. \ref{fig-phogravloop}) can be constructed using our result
\eqref{1photonspinTfin} for the one-photon amplitude in the field, replacing $k$ by $k'$, $\varepsilon$
by the $\upsilon$ of \eqref{defupsilon}, and integrating over $k'$. This gives
\begin{eqnarray}
\Gamma_{\rm spin}^{\rm (tadpole)}(k_0,\epsilon_0;k,\varepsilon;F) &=& 
2ie\kappa
\int_{0}^{\infty}} \! \! dT
{(4\pi T)^{-{D\over 2}}
{\rm e}^{-m^2T}
{\rm det}^{{1\over 2}} \biggl[\frac{\cZ}{\tan\cZ}\biggr]
\int d^D k' \frac{\delta^D (k')}{k'^2}
\nonumber\\ 
&&\hspace{-100pt} \times
k'\cdot
\Bigl({\cot}{\cZ} + {\tan}{\cZ}  - \frac{1}{\cZ}\Bigr)
\cdot \lbrace \epsilon_0,f\rbrace \cdot  k' 
\,{\rm exp}\biggl[\frac{T}{4} 
k'\cdot 
\Bigl(\frac{{\cot}\cZ}{\cZ}- \frac{1}{{\cZ}^2}\Bigr)
\cdot  k'\biggr]
\label{17-tadpole}
\end{eqnarray}
(we have also used the antisymmetry of the Lorentz matrix ${\cot}{\cZ} + {\tan}{\cZ}  - \frac{1}{\cZ}$). 
In the presence of the delta function, the factor ${\rm exp}[k'\cdots k']$ can be replaced by unity since the prefactor
already contains two powers of $k'$. Applying the integral formula \eqref{intk}
leads to
\begin{eqnarray}
\Gamma_{\rm spin}^{\rm (tadpole)}(k_0=-k,\epsilon_0;k,\varepsilon;F) &=& 
\frac{4}{D}ie\kappa
\int_{0}^{\infty}}dT
{(4\pi T)^{-{D\over 2}}
{\rm e}^{-m^2T}
\nonumber\\ 
&&\hspace{-75pt} \times
{\rm det}^{{1\over 2}} \biggl[\frac{\cZ}{\tan\cZ}\biggr]
\tr\biggl\lbrack
\Bigl({\cot}{\cZ} + {\tan}{\cZ}  - \frac{1}{\cZ}\Bigr)
\cdot \epsilon_0 \cdot f \biggr\rbrack \;. 
\label{17-tadpole2}
\end{eqnarray}
This contribution, too, for $D=4$, contains a UV divergence, stemming from the terms linear in the field. 
Thus to calculate it we can replace the determinant factor with unity, and under the trace make the replacement
\be
{\cot}{\cZ} + {\tan}{\cZ}  - \frac{1}{\cZ} \approx \frac{2}{3} {\cZ}  \;.
\ee
Extracting the pole in dimensional regularization we get
\bear
\Gamma_{\rm spin, div}^{({\rm tadpole})}(k_0=-k,\epsilon_0;k,\varepsilon;F)
&=& - \frac{4}{3}
{ie^2\kappa\over(4\pi)^2}{1\over D-4}
\tr (F\epsilon_0 f) \;.
\label{17-tadpolediv}
\ear
As usual, we remove it by a subtraction at the integrand level, and the renormalized tadpole contribution becomes
\begin{eqnarray}
\hspace{-20pt}
\bar\Gamma_{\rm spin}^{\rm (tadpole)}(k_0=-k,\epsilon_0;k,\varepsilon;F) &=& 
\frac{ie\kappa}{16\pi^2}
\int_{0}^{\infty}\frac{dT}{T^2}\, {\rm e}^{-m^2T}
\nonumber\\ 
&& \hspace{-130pt}\times
\biggl\lbrace
{\rm det}^{{1\over 2}} \biggl[\frac{\cZ}{\tan\cZ}\biggr]
\tr\biggl\lbrack
\Bigl({\cot}{\cZ} + {\tan}{\cZ}  - \frac{1}{\cZ}\Bigr)
\cdot \epsilon_0 \cdot f \biggr\rbrack
-  \frac{2}{3}\tr( {\cZ} \epsilon_0 f)
\biggr\rbrace \;.
\label{17-tadpoleren}
\end{eqnarray}
Specializing to a constant magnetic background pointing along the $z$-axis (see Appendix~\ref{app-greendet}), we rewrite the renormalized expression in $D$ dimensions in order to perform the proper-time integral within dimensional regularization. Now the renormalized tadpole contribution is given by 

\bear
\bar{\Gamma}^{(\text{tadpole})}_{\text{spin}}(k_0=-k,\epsilon_0;k,\varepsilon;B)&=&-\frac{ie\kappa}{D}(2\pi)^{-\frac{D}{2}}\,\text{tr}(r_+\epsilon_0 f)\int_0^\infty dT T^{-\frac{D}{2}}\,{\text{e}}^{-m^2T}\non\\
&&\times \bigg\{\frac{z}{3}+ {\coth}\, z-z~ {\coth}^2\,z\bigg\}\non\\
\label{tadpole-spin-T}
\ear
with $z=eBT$ and  ${\cal Z}=zr_+ $. The $z$-integral can be carried out in closed form, leading to an expression of the renormalized tadpole contribution in terms of the Hurwitz zeta function $\zeta(s,a)$,
\bear
\bar\Gamma_{\rm spin}^{\rm (tadpole)}(k_0=-k,\epsilon_0;k,\varepsilon;B)
&=&
\frac{ie\kappa}{DeB}\Big(\frac{\pi}{eB}\Big)^{-\frac{D}{2}} 
\Gamma\Big(1-\frac{D}{2}\Big)\,
\text{tr}(r_+ \epsilon_0 f)\non\\
&&\hspace{-2cm}\times\Bigg[
-\,2^{-\frac{D}{2}}\Big(\frac{m^2}{eB}\Big)^{\frac{D}{2}-1}
-\frac{D}{2}\,\zeta\Big(1-\frac{D}{2},1+\frac{m^2}{2eB}\Big)\non\\
&&\hspace{-1.5cm}+\Big(\frac{D}{2}-1\Big)\frac{m^2}{2eB}\,
\zeta\Big(2-\frac{D}{2},1+\frac{m^2}{2eB}\Big)
+\frac{2^{1-\frac{D}{2}}}{3}\,\Big(1-\frac{D}{2}\Big)\Big(\frac{m^2}{eB}\Big)^{\frac{D}{2}-2}
\Bigg].\non\\
\ear

\begin{figure}[ht]
\begin{center}
\includegraphics[width=0.25\textwidth]{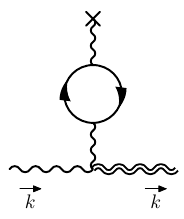}
\caption{Tadpole diagram expanded to linear order in the field.}
\label{fig-tadpoleren}
\end{center}
\end{figure}

However, note that, unlike the renormalization which we performed on the irreducible contribution $\Gamma_{\rm spin}^{\rm (irr)}$,
this one is nothing new; the contribution of the tadpole linear in the field corresponds to the diagram
shown in Fig. \ref{fig-tadpoleren}, which makes it clear that this subtraction is just a special case 
(the zero-momentum limit) of the QED photon wave
function renormalization, 

Note that, differently from the irreducible contribution, for the tadpole contribution the renormalization completely removes the part
linear in the field, leaving only terms of cubic or higher order in the field. 

Let us also give the corresponding result for the scalar loop:

\begin{eqnarray}
\hspace{-20pt}
\bar\Gamma_{\rm scal}^{\rm (tadpole)}(k_0=-k,\epsilon_0;k,\varepsilon;F) &=& 
-\frac{ie\kappa}{32\pi^2}
\int_{0}^{\infty}\frac{dT}{T^2}\, {\rm e}^{-m^2T}
\nonumber\\ 
&& \hspace{-130pt}\times
\biggl\lbrace
{\rm det}^{{1\over 2}} \biggl[\frac{\cZ}{\sin\cZ}\biggr]
\tr\biggl\lbrack
\Bigl({\cot}{\cZ} - \frac{1}{\cZ}\Bigr)
\cdot \epsilon_0 \cdot f \biggr\rbrack
+  \frac{1}{3}\tr( {\cZ} \epsilon_0 f)
\biggr\rbrace \;.
\label{17-tadpolerenscal}
\end{eqnarray}
Analogously to the spinor loop, one obtains the following representation for the scalar case:
\bear
\bar\Gamma_{\rm scal}^{\rm (tadpole)}(k_0=-k,\epsilon_0;k,\varepsilon;F) 
&=&-\frac{2ie\kappa}{DeB}\Big(\frac{2\pi}{eB}\Big)^{-\frac{D}{2}}
\Gamma\Big(1-\frac{D}{2}\Big)\,
\text{tr}(r_+ \epsilon_0 f)\non\\
&&\hspace{-1cm}\times\Bigg[
\frac{D}{2}\zeta\Big(1-\frac{D}{2},\half+\frac{m^2}{2eB}\Big)
+\Big(1-\frac{D}{2}\Big)\frac{m^2}{2eB}\zeta\Big(2-\frac{D}{2},\half+\frac{m^2}{2eB}\Big)\non\\
&&
+\frac{2^{-\frac{D}{2}}}{3}\Big(1-\frac{D}{2}\Big)\Big(\frac{m^2}{eB}\Big)^{\frac{D}{2}-2}
\Bigg].
\ear

\section{Third contribution to electromagnetic photon-graviton conversion}
\label{reducible}
\renewcommand{\theequation}{6.\arabic{equation}}
\setcounter{equation}{0}

For the other reducible contribution, corresponding to the third diagram in Fig. \ref{fig-phogravloop}, we can use 
eqs. \eqref{17-Gammaext}, \eqref{defupsilonF}, with $k'=k_0=-k$, and our results for the vacuum polarization tensor in a constant field
from subsection \ref{subsec:The vacuum polarization tensor in a constant field}. 
Like the tadpole contribution, it has a UV divergence from the field-independent part of the vacuum polarization tensor. 
This part corresponds to the right-hand diagram of Fig. \ref{fig-phogravloop}, but with the loop fermion taken in vacuum,
and according to the LSZ formalism is not to be taken into account in the calculation of the on-shell amplitude. 
Using our result \eqref{vpspinren} for the on-shell renormalized one-loop vacuum polarization tensor in the constant field, and
putting things together, we have
\bear
\bar\Gamma_{\rm spin}^{\rm (ext)}(k_0=-k,\epsilon_0;k,\varepsilon;F) 
&=& v_{F\beta}\Bigl(\bar\Pi_{\rm spin}^{\beta\alpha}(k;F)-\bar\Pi_{\rm spin}^{\beta\alpha}(k;0)\Bigr)
\varepsilon_{\alpha}\, ,
\label{Gammaext}
\ear
with
\bear
\upsilon_F 
 = i \kappa \frac{\lbrace \epsilon_0,F\rbrace\cdot k_0}{k_0^2}
= -i \kappa \frac{\lbrace \epsilon_0,F\rbrace\cdot k}{k^2}
\, ,
\label{17-upsilonF}
\ear
\bear
\bar\Pi^{\beta\alpha}_{\rm spin}(k;F)
&=&
\frac{e^2}{8\pi^2}
{\dps\int_{0}^{\infty}}{dT\over T}
{\rm e}^{-m^2T}
\int_0^1 du_1
\biggl\lbrace
\dettan
\, \tilde I^{\beta\alpha}_{\rm spin}
\,\e^{-Tk\cdot\Phi_{12}\cdot k}
\nonumber\\&&
\hspace{120pt}
 + \frac{2}{3} \bigl(\delta^{\beta\alpha}k^2 -k^{\beta}k^{\alpha}\bigr)
\biggr\rbrace\,,
\label{17-vpspinren}
\ear\no
and $\tilde I^{\beta\alpha}_{\rm spin}$ as given in \eqref{intfinalspin}. 
Note that this contribution, too, after renormalization starts with terms cubic in $F$, and that on-shell it has an IR divergence
through the pole in \eqref{17-upsilonF}. 
A proper treatment of the on-shell limit requires a somewhat different interpretation of the third diagram,
namely as a correction to the tree-level amplitude due to the modified dispersion relation of the photon 
in the field \cite{ahjari}. In the presence of an external field, the physical photon eigenmodes no longer satisfy the vacuum on-shell condition $k^2=0$; instead, their propagation is governed by the dispersion relation $k^2+\Pi(k)=0$, where $\Pi$ denotes the photon polarization tensor, leading in general to a field induced deviation from $k^2=0$. 

The corresponding amplitude for a scalar loop is simply obtained by replacing $\bar\Pi^{\beta\alpha}_{\rm spin}(k;F)$ with $\bar\Pi^{\beta\alpha}_{\rm scal}(k;F)$ in \eqref{Gammaext}.

\section{Weak-field expansion of the tadpole contribution} 
\label{weak-B}

Since phenomenological applications are primarily concerned with the regime of weak external fields, we briefly discuss the tadpole contribution in the limit of a weak magnetic background,
\begin{equation}
eB \ll m^2 \, .
\end{equation}
In this regime, the proper-time integrals admit a systematic asymptotic expansion in powers of the dimensionless ratio $eB/m^2$.

For definiteness, we present the weak-field expansion only for the spinor loop, since the scalar case follows by an entirely analogous procedure. To derive this expansion, it is convenient to return to the proper-time representation of the spinor tadpole contribution given in Eq.~(\ref{tadpole-spin-T}), prior to performing the $z$-integration. Specializing to $D=4$, the hyperbolic functions appearing in the integrand can be expanded for small
\begin{equation}
z = eBT \, ,
\end{equation}
which is justified since the proper-time integral is exponentially dominated by values $T\sim 1/m^2$ in the weak-field regime.

A compact representation of the weak-field expansion follows from the series representation of the integrand in Eq.~(\ref{tadpole-spin-T}),
\begin{equation}
\frac{z}{3}+{\coth}\, z - z\,{\coth^2}\, z
=
\sum_{n=2}^{\infty}
\frac{2^{2n}\mathcal{B}_{2n}}{(2n-1)!}\,z^{2n-1},
\end{equation}
where $\mathcal{B}_{n}$  denotes the Bernoulli numbers. Substituting this expansion into the proper-time integral and carrying out the integration term by term yields the following closed-form weak-field expansion for the renormalized spinor tadpole contribution:
\begin{eqnarray}
\bar{\Gamma}^{(\text{tadpole})}_{\text{spin}}(k_0=-k,\epsilon_0;k,\varepsilon;B)
&=&
-\frac{ie\kappa}{4eB}(2\pi)^{-2}(eB)^2\,
\text{tr}(r_+\epsilon_0 f)
\nonumber\\
&&\times
\sum_{n=2}^{\infty}
\frac{2^{2n}\mathcal{B}_{2n}}{(2n-1)!}
\Big(\frac{eB}{m^2}\Big)^{2n-2}
\Gamma(2n-2)\,.
\end{eqnarray}
The corresponding weak-field expansion for the scalar loop can be obtained by the same method and will not be displayed here.

\section{Ward identities for the photon-graviton amplitude in a constant field}
\label{ward}
\renewcommand{\theequation}{7.\arabic{equation}}
\setcounter{equation}{0}

Finally, let us discuss how the Ward identities work for the photon-graviton amplitude in a constant field,
returning to the off-shell unrenormalized amplitudes. Here there is no difference between the scalar and spinor-loop cases.
It is easy to see that the gauge
Ward identity holds for each of the three contributions separately, i.e.
\bear
\Gamma^{\rm (irr),(tadpole), (ext)}(-k,\epsilon_0;k,\delta\varepsilon;F) = 0 \,,
\label{17-wardgauge}
\ear
where $\delta\varepsilon_i^\mu = k^\mu$.  

For the gravitational Ward identity, defined by the transformation
\bear
\delta\epsilon_0^\mn &=& k_0^\mu\zeta_0^\nu + \zeta_0^\mu k_0^\nu \,,
\ear
a slight extension of the manipulations of the appendix of \cite{61}
shows that the inclusion of the one-photon amplitude leads to
\bear
\Gamma^{\rm (irr)}(k_0,\delta\epsilon_0;k,\varepsilon;F) =
\Gamma(k_0+k,\tilde\varepsilon;F) 
+
\Gamma(k_0,\tilde\varepsilon_F;k,\varepsilon;F)\,,
\label{17-wardphograv}
\ear
where on the right-hand side we have now the one- and two-photon amplitudes, and
\bear
\tilde \varepsilon_i^\mu \equiv \kappa  (k_i^{\mu}\varepsilon_i^{\nu} - \varepsilon_i^{\mu}k_i^{\nu}) 
\zeta_{0\nu}   = \kappa (f_i\cdot \zeta_0)^\mu\,, \nonumber\\
\tilde\epsilon_F^\mu \equiv -i\kappa(F\cdot\zeta_0)^\mu \;.
\label{deftildeepsilonF}
\ear
However, the first term on the right-hand side of \eqref{17-wardphograv} vanishes when setting $k_0=-k$. 
Using
\begin{equation}
\delta\upsilon_F^{\mu} =  -\tilde \varepsilon_F^{\mu} -i 
\kappa\frac{k_0\cdot F \cdot \zeta_0}{k_0^2} k_0^{\mu} \, ,
\label{17-deltaupsilonF}
\end{equation}
we can get from \eqref{17-wardphograv} the following on-shell
Ward identity connecting $\Gamma_{\rm spin}^{\rm (irr)}$ and $\Gamma_{\rm spin}^{\rm (ext)}$, 
\bear
\Gamma^{\rm (irr)}(-k,\delta\epsilon_0;k,\varepsilon;F)
+
\Gamma^{\rm (ext)}(-k,\delta\epsilon_0;k,\varepsilon;F)
=0 \;,
\label{17-wardphogravos}
\ear
while the tadpole contribution on-shell becomes invariant by itself,
\bear
\Gamma^{\rm (tadpole)}(-k,\delta\epsilon_0;k,\varepsilon;F) =0\,,
\ear
as can be easily seen from \eqref{17-tadpole2} using the on-shell relations $k^2=\varepsilon\cdot k = \zeta_0\cdot k_0 =0$.

\section{Conclusions}
\label{conc}
In this work we have re-examined electromagnetic photon–graviton conversion in a constant external field within Einstein–Maxwell theory at the one-loop level. Although photon–graviton conversion in an electromagnetic field might eventually find applications in a broader range of physical contexts, here we have focused on its dichroism aspect, motivated by the analysis of Ref.~\cite{ahjari}, which demonstrated that this process provides the leading Standard Model contribution to magnetic dichroism, despite its small absolute magnitude under realistic conditions.

Our main result is the identification and explicit evaluation of a non-vanishing tadpole contribution to the one-loop photon–graviton conversion amplitude, corresponding to the middle diagram in Fig.~\ref{fig-phogravloop}. Contrary to long-standing assumptions, this diagram does not vanish once the infrared behavior of the connecting photon propagator is treated correctly in dimensional regularization. Using the worldline formalism, we have provided a unified calculation of all one-loop contributions — irreducible, tadpole, and other reducible — for scalar and spinor loops in a general constant electromagnetic background.

Although the tadpole diagram contributes to the full one-loop amplitude, it is irrelevant for magnetic dichroism already at a structural level.
From \eqref{tadpole-spin-T} we see that the magnetic tadpole contribution depends on the photon and graviton polarizations through a global factor
${\rm tr}(F\epsilon_0 f)$, which (for on-shell gravitons) is proportional to the tree-level amplitude
$\Gamma^{\rm (tree)}(k_0,\epsilon_0;k,\varepsilon;F)$ of eq. \eqref{phogravtree},
implying identical photon–graviton conversion rates for the two photon polarizations. Consequently, the magnetic dichroism remains unchanged, and the analysis of Ahlers \textit{et al.}~\cite{ahjari} is not affected by the presence of the tadpole diagram.

In the weak-field regime $eB/m^2\ll1$, relevant for phenomenological applications, the tadpole contribution is moreover parametrically suppressed compared to the irreducible one-loop contribution. In the strong-field limit $eB/m^2\gg1$, the relative importance of different one-loop contributions can change qualitatively, as has been demonstrated in related contexts \cite{karbstein-19}. In such regimes, tadpole-type contributions may play a more prominent role in the overall amplitude, which could be relevant not only from a theoretical standpoint but also in extensions of the Standard Model involving light or weakly charged particles, such as mini-charged particles \cite{doebrich-12}, for which strong-field effects can become physically relevant.

Independently of these phenomenological considerations, the inclusion of the tadpole diagram is conceptually important. It completes the one-loop description of photon–graviton conversion in external fields and clarifies its close connection to other one-particle reducible effects in background-field QED. Our results thus provide a consistent and complete framework for future studies, including strong-field backgrounds, non-standard charged sectors, and higher-loop corrections.

\appendix
\section{Einstein-Maxwell theory}
\label{appA}
\renewcommand{\theequation}{A.\arabic{equation}}
\setcounter{equation}{0}

In our euclidean conventions, the action for a scalar field coupled to electromagnetism and gravity is
\bea
S[\phi,\phi^* ;g,A] = \int d^Dx \sqrt{g}
\Big [g^{\mu\nu} (\partial_\mu -i e A_\mu) \phi^* (\partial_\nu + ie A_\nu)
\phi + (m^2 +\xi R) \phi^* \phi \Big ] \;.
\label{16SgA}
\eea
In the spin 1/2 case, the euclidean action for a Dirac field $\Psi$ coupled to
electromagnetism and gravity is given by
\bea
S[\Psi,\bar \Psi;e, A]
= \int \!d^Dx \; \sqrt{g}\, \bar{\Psi} (\rldd + m)\Psi  
\eea
where the vielbein $e_\mu^a$ and the spin connection $\omega_{\mu ab} $ appear inside the Dirac operator
\be
\rldd = \gamma^a e_a^\mu \nabla_{\!\mu} \ , \quad \quad 
\nabla_{\!\mu} = \partial_\mu + i e A_\mu 
+ {1\over 4}\omega_{\mu ab} \gamma^a \gamma^b \;.
\ee
Moreover, we will need also the interaction between photons and gravitons, which is described by the 
action of Einstein-Maxwell theory (in euclidean conventions)
\bea  
S[g,A] =    
\int d^D x\ \sqrt{g}\, \bigg (  
-{1\over 2 \kappa^2 } R + {1\over 4}F_{\mu\nu}F^{\mu\nu}  
\bigg )   
\label{EM}  
\eea  
where $\kappa^2 = 8\pi G_N$.  
Let us expand this action to quadratic order in the fields. 
Expanding $g_{\mu\nu} = \delta_{\mu\nu}+\kappa h_{\mu\nu}$ and $A_\mu =\bar A_\mu +a_\mu$,
so that $F_{\mu\nu}= \bar F_{\mu\nu} + f_{\mu\nu}$,
and using the short-hand notation 
$h\equiv \delta^{\mu\nu} h_{\mu\nu}$, one obtains 
the following quadratic approximation in the fluctuations $(h_{\mu\nu}, a_\mu)$ around the background $(\delta_{\mu\nu}, \bar A_\mu)$:
\begin{align}
S_{(2)} &=  
\int d^D x \, \bigg \{  \frac14 (\pa_\alpha h_{\mu\nu})^2 -\frac18  (\pa_\alpha h)^2 
- \frac12 \bigg (  \pa^\alpha  h_{\alpha \mu} - \frac12 \pa_\mu h \bigg)^2 
+ \frac12 (\pa_\mu a_\nu)^2 - \frac12 (\pa^\mu a_\mu)^2 
\cr
 &-  \frac{\kappa}{2} h_{\mu\nu}\Big (\bar F^{\mu\alpha}\bar F^\nu{}_\alpha  
- \frac14 \delta^{\mu\nu} \bar F^2 \Big)  
\cr
&-   
\kappa h_{\mu\nu}\Big (   
\bar F^{\mu\alpha} f^\nu{}_\alpha  
- {1\over 4} \delta^{\mu\nu} \bar F^{\alpha\beta} f_{\alpha\beta}   
\Big )  
\cr
&+    {\kappa^2\over 4} \Big [   
\Big ( {1\over 8} h^2 -{1\over 4} h_{\mu\nu}^2 \Big )\bar F^2  
+ h^{\mu\nu} h^{\alpha\beta}  
\bar F_{\mu\alpha}\bar F_{\nu\beta}   
+ (2 h^{\mu \alpha} h^\nu{}_\alpha -h h^{\mu\nu} )  
\bar F_{\mu\beta}\bar F_\nu{}^\beta\Big ]\bigg \}  \;.
\label{17-2}
\end{align}

In the second line of this expression, we recognize the linear coupling 
$-{\kappa\over 2} h_{\mu\nu} \bar T^{\mu\nu}$ of the graviton $h_{\mu\nu}$ 
to the stress tensor of the background electromagnetic field
\bear
 \bar T^{\mu\nu}=\bar F^{\mu\alpha}\bar F^\nu{}_\alpha  - {1\over 4} \delta^{\mu\nu} \bar F^2  \;.
 \ear
The existence of this tadpole vertex indicates that the nontrivial background stress tensor tends to curve space.
The third line in (\ref{17-2}) gives instead the tree-level graviton-photon conversion
in the electromagnetic background.
Using plane waves 
\be
h_{\mn}(x) = \ep_{0\mn}\,  {\rm e}^{ik_0x} \ , \qquad
a_{\alpha}(x) = \varepsilon_{\alpha}\,  {\rm e}^{ik x} 
\ee
we get for this conversion term the vertex (deleting now the `bar' on $F$)
\bea
\Delta S_{(2)}  
= (2\pi)^D \delta^D(k_0+k)\, \ep_{0\mn}  \varepsilon_{\alpha}  \, 
(-i \kappa) \,
\Big [ F^{\mu\alpha}k^\nu - (F\cdot k)^\mu \delta^{\nu\alpha}
+{1\over 2} \delta^{\mn} (F\cdot k)^\alpha \Big ] \;.
\nonumber
\eea

The two-point functions (which we denote by $\Pi$),
in either coordinate or momentum space, 
are contained in $S_{(2)}$ (or in the quadratic part of
the full effective action 
$\Gamma_{(2)}   = - S_{(2)}  + \Gamma_{(2)}^{{\rm (1-loop)}}  + \cdots $)  
as follows
\footnote{We Fourier transform fields as 
$h_{\mu\nu}(k)  = \int d^D x\, {\rm e}^{-ikx} \, h_{\mu\nu}(x) $, 
with inverse Fourier transform given by
$h_{\mu\nu}(x)  = \int \frac{d^{D} k}{(2\pi)^D} h_{\mu\nu}(k)\, {\rm e}^{ikx} $. }
 \bea  
S_{(2)} 
 \eqa  
\int d^D x \, \bigg \{ 
{1\over 2}  h_{\mu\nu}(x) \Pi^{\mn,\lambda\rho}(-i\partial) h_{\lambda\rho}(x) 
+ {1\over 2} a_\alpha(x) \Pi^{\alpha\beta} (-i\partial) a_{\beta}(x) \ccr
&& +  h_{\mu\nu}(x) \Pi^{\mn,\alpha}(-i\partial) a_\alpha(x) 
- {\kappa \over 2}  h_{\mu\nu}(x) \bar T^{\mu\nu}(x) \bigg \}\ccr
\eqa  
\int {d^D k\over (2 \pi)^{D}} \, \bigg \{  
 {1\over 2}h_{\mu\nu}(-k) \Pi^{\mn,\lambda\rho}(k) h_{\lambda\rho}(k) 
+ {1\over 2} a_\alpha(-k) \Pi^{\alpha\beta} (k) a_{\beta}(k) \ccr
&& +  h_{\mu\nu}(-k) \Pi^{\mn,\alpha}(k) a_\alpha(k) 
- {\kappa\over 2}  h_{\mu\nu}(-k) \bar T^{\mu\nu}(k) \bigg \} \;.
\label{6}
\eea
The equations of motion in terms of these two-point functions then
read
\bear
\delta  a_{\alpha}(-k) : \quad &&  
\Pi^{\alpha\beta} (k) a_{\beta}(k) + \Pi^{\mn,\alpha}(-k) h_{\mu\nu}(k) =0
\label{eoma}
\\
\delta  h_{\mu\nu}(-k) : \quad  && 
\Pi^{\mn,\lambda\rho}(k) h_{\lambda\rho}(k)+
 \Pi^{\mn,\alpha}(k) a_\alpha(k) = {\kappa\over 2} \bar T^{\mu\nu}(k) 
\label{eom}
\ear
and, in particular, one obtains from \eqref{17-2}
\bea
\Pi^{\alpha\beta}_{\rm tree} (k) &=& k^2 \delta^{\alpha\beta} - k^\alpha k^\beta \, ,
\label{17-Piphoton}\\
\Pi^{\mn,\alpha}_{\rm tree} (k)  &=& -{i \kappa\over 2} C^{\mn,\alpha} \label{Pitree}
\eea
with 
\be
C^{\mn,\alpha} =
F^{\mu\alpha}k^{\nu} +F^{\nu\alpha}k^{\mu}
-\bigl(F\cdot k\bigr)^{\mu}\delta^{\nu\alpha}
-\bigl(F\cdot k\bigr)^{\nu}\delta^{\mu\alpha} 
+ \bigl(F\cdot k\bigr)^{\alpha}\delta^{\mn} 
\label{app-defCmna}
\ee
depicted in Fig. \ref{fig-3vertex}.

From the cubic fluctuations, we will still need the term coupling the graviton to two photons:
\bear
S_{(3)}^{G\gamma\gamma}
=- {\kappa\over 2} \int d^Dx \,  h_{\mu\nu}\Big (    
f^{\mu\alpha}f^\nu{}_\alpha  
- {1\over 4} \delta^{\mu\nu} f^2
\Big)  \;.
\label{S3}
\ear

\section{Worldline Green's functions and determinants in a constant field}
\label{app-greendet}
\renewcommand{\theequation}{B.\arabic{equation}}
\setcounter{equation}{0}

Define ${\cal Z}_{\mu\nu} \equiv eF_{\mu\nu}T\, .$

\subsection{Periodic case}
\label{subsec:appC-FP}

\noindent
{\it Determinant factor:}
\bear
{\rm det}^\half \biggl[\frac{\cZ}{\sin \cZ}\biggr] \, .
\label{app-gd-detP}
\ear

\noindent
{\it Worldline Green's function and its derivatives:}
\begin{eqnarray}
{\cal G}_{B12} &=& \frac{T}{2{\cal Z}^2}
\biggl({{\cal Z}\over{{\rm sin}{\cal Z}}}
\,{\rm e}^{-i{\cal Z}\dot G_{B12}}
+i{\cal Z}\dot G_{B12} -1\biggr)
\, ,
\label{app-gd-calGB}\\
\dot{\cal G}_{B12}
&=&
{i\over {\cal Z}}\biggl({{\cal Z}\over{{\rm sin}{\cal Z}}}
\,{\rm e}^{-i{\cal Z}\dot G_{B12}}-1\biggr)
\, ,
\label{app-gd-dotcalGB}\\
\ddot{\cal G}_{B12}
&=& 2\delta_{12} -{2\over T}{{\cal Z}\over{{\rm sin}{\cal Z}}}
\,{\rm e}^{-i{\cal Z}\dot G_{B12}}
\label{app-gd-ddotcalGB}
\end{eqnarray}

\noindent
{\it Symmetry properties:}
\bear
{\cal G}_{B12} = 
{\cal G}_{B21}^{T},
\quad
\dot{\cal G}_{B12} = 
-\dot{\cal G}_{B21}^{T},
\quad
\ddot{\cal G}_{B12} = 
\ddot{\cal G}_{B21}^{T}.
\label{app-gd-symmcalGB}
\ear\no

\noindent
{\it Coincidence limits:}
\bear
{\cal G}_{B}(\tau,\tau) &=&
{T\over 2}
\Bigl(
\frac{\cot{\cal Z}}{\cal Z} 
- \frac{1}{{\cal Z}^2}
\Bigr)
\, ,
\label{app-gd-coincalGB}\\
\dot {\cal G}_B(\tau,\tau) &=& i{\rm cot}{\cal Z}
-{i\over {\cal Z}}
\, ,
\label{app-gd-coindcalGB}\\
\ddot {\cal G}_B(\tau,\tau) &=& 
2\delta(0) -\frac{2}{T}{\cal Z}\cot \cZ  \;.
\label{app-gd-coinddcalGB}
\end{eqnarray}
\noindent
{\it Weak-field expansion:}

\begin{eqnarray}
{\cal G}_{B12} &=& {G}_{B12}-{1\over 6}
-{i\over 3}
\dot G_{B12}{G}_{B12}{\cal Z} + \Bigl({1\over 3T}{G}_{B12}^2
-{T\over 90}\Bigr){\cal Z}^2+O({\cal Z}^3) \, , \label{app-gd-GBexpand}\\
\dot{\cal G}_{B12} 
&=&\dot G_{B12}+2i\Bigl(\frac{G_{B12}}{T}-{1\over 6}\Bigr){\cal Z}
+{2\over 3T}\dot G_{B12}{G}_{B12}{\cal Z}^2 +  O({\cal Z}^3) 
\, ,
\label{app-gd-dGBexpand}\\
\ddot{\cal G}_{B12} 
&=& 2 \delta_{12} - \frac{2}{T} +\frac{2}{T}i\dot G_{B12}{\cal Z}
-4\Bigl(\frac{G_{B12}}{T}-{1\over 6}\Bigr){\cal Z}^2+O({\cal Z}^3)  \;.
\label{app-gd-ddGBexpand}
\end{eqnarray}

\medskip

\noindent
{\it Even-odd decomposition:}

For the purpose of simplifying integrals, 
it will often be convenient to decompose the constant-field worldline Green's functions as
\bear
{\cal G}_{B12} = {\cal S}_{B12} + {\cal A}_{B12}
\label{app-gd-evenoddcalGB}
\ear
where ${\cal S}_{B}$ denotes the even and ${\cal A}_{B}$ the odd part as a power series in $F$.
Note that ${\cal S}_B$ is symmetric as a Lorentz matrix and ${\cal A}_{B}$ antisymmetric, which motivates
our notation. 
%
%
\medskip

\noindent
{\bf Special constant fields:}

\begin{itemize}

\item
{\bf Parallel electric and magnetic fields:} 

\noindent
in the generic case, it is possible to find a Lorentz system where 
$\bf E$ and $\bf B$ both point along the $z$ - axis.
If we further assume, for definiteness, that ${\bf E}\cdot{\bf B} > 0$, then we can 
take them both to point along the positive $z$-axis, 
${\bf E} = (0,0,E), {\bf B} = (0,0,B)$.
The euclidean field strength tensor then takes the form
\begin{equation}
F =
\left(
\begin{array}{*{4}{c}}
0&B&0&0\\
-B&0&0&0\\
0&0&0&iE\\
0&0&-iE&0
\end{array}
\right) \;.
\label{app-gd-Fspecial}
\end{equation}
\vspace{10pt}

\noindent
Defining $z_+ \equiv eBT$ and $z_- \equiv ieET$, 
the determinant factor \eqref{app-gd-detP} can then be written as
\bear
{\rm det}^{-{1\over 2}}
\biggl[
\frac{\sin\cZ}{\cZ}
\biggr] &=&
\frac{z_+z_-}{\sinh z_+\sinh z_-}
=
\frac{eBT}{\sinh eBT}\frac{eET}{\sin eET}
\, .
\label{app-gd-detP-special}
\ear
We further define Lorentz matrices $g_+$, $g_-$, and $r_+$, $r_-$ by
\bear
g_+\equiv
\left(
\begin{array}{*{4}{c}}
1&0&0&0\\
0&1&0&0\\
0&0&0&0\\
0&0&0&0
\end{array}
\right),\qquad
g_-\equiv
\left(
\begin{array}{*{4}{c}}
0&0&0&0\\
0&0&0&0\\
0&0&1&0\\
0&0&0&1
\end{array}
\right),\nonumber\\
\label{app-gd-gmat}
\ear
\vspace{-30pt}
\begin{equation}
r_+ \equiv
\left(
\begin{array}{*{4}{c}}
0&1&0&0\\
-1&0&0&0\\
0&0&0&0\\
0&0&0&0
\end{array}
\right),\qquad
r_- \equiv
\left(
\begin{array}{*{4}{c}}
0&0&0&0\\
0&0&0&0\\
0&0&0&1\\
0&0&-1&0
\end{array}
\right).
\nonumber
\end{equation}
Then one can show the following matrix decompositions of the worldline Green's functions 
%
\bear
{\cal S}_{B12}^{\mu\nu}
&=&
-{T\over 2}
\sum_{\alpha =\pm }
\frac{A_{B12}(z_{\alpha})}{z_{\alpha}}\,g_{\alpha}^{\mu\nu}
\, ,
\label{app-gd-specialdecompcalSAfirst}
\\
{\cal A}_{B12}^{\mu\nu}
&=&
{iT\over 2}
\sum_{\alpha =\pm }
{S_{B12}(z_{\alpha})-\dot G_{B12}\over z_{\alpha}}
\,r_{\alpha}^{\mu\nu}
\, ,
\\
\dot{\cal S}_{B12}^{\mu\nu} &=&
\sum_{\alpha =\pm }
S_{B12}(z_{\alpha})\,g_{\alpha}^{\mu\nu}
\, ,
\\
\dot{\cal A}_{B12}^{\mu\nu} &=& 
-i
\sum_{\alpha =\pm }
A_{B12}(z_{\alpha})\,r_{\alpha}^{\mu\nu}
\, ,
\\
\ddot{\cal S}_{B12}^{\mu\nu} &=& \ddot G_{B12}\delta^{\mu\nu}
-{2\over T}
\sum_{\alpha =\pm }
z_{\alpha}A_{B12}(z_{\alpha})\,g_{\alpha}^{\mu\nu}
\, ,
\\
\ddot{\cal A}_{B12}^{\mu\nu} &=& 
{2i\over T}
\sum_{\alpha =\pm }
z_{\alpha}S_{B12}(z_{\alpha})\,r_{\alpha}^{\mu\nu} \;.
\label{app-gd-specialdecompcalSAlast}
\ear\no
These decompositions involve essentially only the two scalar, dimensionless functions
$S_{B12}$ and $A_{B12}$,
\bear
S_{B12}(z) &=&
{\sinh(z\,\dot G_{B12})\over \sinh z} \, ,
\label{defS}\\
A_{B12}(z) &=&
{\cosh(z \,\dot G_{B12})\over 
\sinh z}-{1\over z} \;.
\label{app-defA}
\ear
Of those only $A_{B12}$ has a non-vanishing coincidence limit,
\bear
A_{Bii} &=&
{\rm coth} \, z -\frac{1}{z} \;.
\label{app-gd-coinAB}
\ear\no
Let us also write down the first few terms of the weak field
expansions of these functions,
\bear
S_{B12}(z) &=&
\dot G_{B12}\biggl[
1-{2\over 3}{G_{B12}\over T}z^2
+\Bigl({2\over 45}{G_{B12}\over T}+
{2\over 15}{G_{B12}^2\over T^2}\Bigr)z^4
+\, {\rm O}(z^6)
\biggr]
\, ,
\\
A_{B12}(z) &=&
\Bigl(\third -2{G_{B12}\over T}\Bigr)z
+\Bigl(-{1\over 45}+{2\over 3}
{G_{B12}^2\over T^2}\Bigr)z^3
+\,  {\rm O}(z^5) \;.
\ear\no

\item
{\bf The purely magnetic field case:}

We let the magnetic field point along the positive $z$-axis and denote $z=eBT$.
Then the determinant factor \eqref{app-gd-detP} becomes
\bear
{\rm det}^{-{1\over 2}}
\biggl[{\sin\cZ\over {{\cal Z}}}
\biggr]&=&
{z\over{\sinh z}} 
\, .
\label{app-gd-detB}
\ear
The decompositions 
\eqref{app-gd-specialdecompcalSAfirst} - \eqref{app-gd-specialdecompcalSAlast} 
now involve only the three matrices $g_+$, $g_-$ and $r_+$:
\begin{eqnarray}
{\cal G}_{B}(\tau_1,\tau_2) 
&=&G_{B12}\,{g_-}
-{T\over 2z}
A_{B12}(z)
{g_+}
+{T\over{2z}}\Bigl(S_{B12}(z)
-\dot G_{B12}\Bigr)i{r_+}\, ,  \quad \\
\dot{\cal G}_{B}(\tau_1,\tau_2)
&=&\dot G_{B12}\,{g_-}+ S_{B12}(z)g_+ -A_{B12}(z) i{r_+}\, , \\
\ddot{\cal G}_{B}(\tau_1,\tau_2)
&=& \ddot G_{B12}\Eins -\frac{2z}{T} A_{B12}(z)g_+
+ \frac{2z}{T} S_{B12}(z) i r_+ \;.
\label{D-Gddmag}
\ear
The coincidence limits \eqref{app-gd-coincalGB} - \eqref{app-gd-coinddcalGB} turn into
\begin{eqnarray}
{\cal G}_B(\tau,\tau) &=& 
-\frac{T}{6}\, {g_-}
-{T\over 2z}
\biggl({\coth }\, z- \frac{1}{z}\biggr)
{g_+}
=
-\frac{T}{6}\Eins -{T\over 2z}
\biggl({\coth} \, z- \frac{1}{z}-\frac{z}{3} \biggr)
{g_+}
\, ,
\nonumber\\
\\
\dot {\cal G}_B(\tau,\tau) &=& - \biggl({\coth }\, z- \frac{1}{z}\biggr)
i{r_+}\, ,\\
\ddot {\cal G}_B(\tau,\tau) &=& 
2\Bigl(\delta (0) - \frac{1}{T}\Bigr)\Eins
-\frac{2z}{T} \biggl({\coth} \, z- \frac{1}{z}\biggr) g_+ \;.
\label{app-gd-coinddcalGmag}
\end{eqnarray}

\item
{\bf The purely electric field case:}

To obtain the analogue of the magnetic formulas \eqref{app-gd-detB} - \eqref{app-gd-coinddcalGmag} for a purely electric field pointing into the $z$-direction, just replace $z=eBT$ by $z=ieET$ and 
interchange $+ \leftrightarrow -$ everywhere. 

\item
{\bf The crossed-field case:}

In a ``crossed field'', defined by 
${\bf E}\perp{\bf B}, E=B$, both invariants 
$B^2-E^2$ and ${\bf E}\cdot{\bf B}$
vanish. For such a field $F^3 =0$, so that the power series 
\eqref{app-gd-GBexpand}-\eqref{app-gd-ddGBexpand}
break off after their quadratic terms.
The worldline correlators thus can be represented by the terms given there. 

\noindent
The determinant factor becomes unity.

\end{itemize}

\subsection{Anti-periodic case}
\label{subsec:appC-FAP}

\noindent
{\it Determinant factor:}
\bear
{\rm det}^\half \bigl[\cos \cZ\bigr] \, .
\label{app-gd-detAP}
\ear

\noindent
{\it Worldline Green's function and its derivative:}
\begin{eqnarray}
{\cal G}_{F}(\tau_1,\tau_2) &=&
G_{F12}
{{\rm e}^{-i{\cal Z}\dot G_{B12}}\over {\rm cos}{\cal Z}} \, ,
\label{app-gd-calGF}\\
\dot {\cal G}_{F}(\tau_1,\tau_2) &=&
2\delta_{12} + \frac{2i}{T}G_{F12}
\frac{\cZ}{\cos\cZ}
{\rm e}^{-i{\cal Z}\dot G_{B12}} \, .
\label{app-gd-dcalFGF}
\end{eqnarray}

\noindent
{\it Symmetry properties:}
\bear
{\cal G}_{F12} = 
-{\cal G}_{F21}^{T},
\quad
\dot{\cal G}_{F12} = 
\dot{\cal G}_{F21}^{T}
\, .
\label{app-gd-symmcalGF}
\ear\no

\noindent
{\it Coincidence limits:}
\bear
{\cal G}_F(\tau,\tau) &=& -i\,{\rm tan}{\cal Z} \, ,
\label{app-gd-coincalGF}
\\
\dot{\cal G}_F(\tau,\tau) &=& 2\delta(0) + \frac{2}{T}{\cal Z}\tan \cZ 
\, .
\label{app-gd-coincaldGF}
\end{eqnarray}

\noindent
{\it Weak-field expansion:}
\begin{eqnarray}
{\cal G}_{F12}&=& G_{F12}-iG_{F12}\dot G_{B12}{\cal Z}
+2G_{F12}\frac{G_{B12}}{T}{\cal Z}^2+O({\cal Z}^3) 
\, ,
\label{app-gd-GFexpand}
\\
\dot{\cal G}_{F12}&=& 2\delta_{12} +\frac{2}{T}i{\cal Z}G_{F12}
+\frac{2}{T}G_{F12}\dot G_{B12}{\cal Z}^2 + O({\cal Z}^3)  \;.
\label{app-gd-dGFexpand}
\end{eqnarray}

\noindent
{\it Even-odd decomposition:}
\bear
{\cal G}_F = {\cal S}_F + {\cal A}_F \, .
\label{app-gd-evenoddcalF}
\ear

\medskip

\noindent
{\bf Special constant fields:}

\begin{itemize}

\item
{\bf Parallel electric and magnetic fields:}

\noindent
With the same notation as in the periodic case above, 
\bear
{\rm det}^{{1\over 2}} \bigl[ \cos\cZ \bigr] 
&=&
\cosh z_+\cosh z_- = \cosh(eBT)\cos(eET)
\, ,
\label{app-gd-detAPmag}
\ear
\bear
{\cal S}_{F12}^{\mu\nu} &=&
\sum_{\alpha =\pm }
S_{F12}(z_{\alpha})\,g_{\alpha}^{\mu\nu}
\, ,
\\
{\cal A}_{F12}^{\mu\nu} &=& 
-
\sum_{\alpha =\pm }
A_{F12}(z_{\alpha})\,ir_{\alpha}^{\mu\nu}
\, ,
\\
\dot {\cal S}_{F12}^\mn &=& \dot G_{F12}\Eins -\frac{2}{T} \sum_{\alpha =\pm }
A_{F12}(z_{\alpha})\,g_{\alpha}^{\mu\nu}
\, ,
\\
\dot {\cal A}_{F12}^\mn &=&
\frac{2}{T}
\sum_{\alpha =\pm }
z_{\alpha}S_{F12}(z_{\alpha})\,ir_{\alpha}^{\mu\nu} \;.
\ear\no
The scalar, dimensionless 
coefficient functions appearing in these formulas are 
\bear
S_{F12}(z) &=&
G_{F12}{\cosh(z\,\dot G_{B12})\over\cosh(z)}
\, ,
\\
A_{F12}(z) &=&
G_{F12}{\sinh(z\,\dot G_{B12})\over \cosh(z)} \;.
\ear
Of those, it is $A_{F12}$ that has a non-vanishing coincidence limit
\bear
A_{Fii} &=&
{\tanh} \, z \;.
\label{app-gd-coinAF}
\ear\no
As for the periodic case above, let us also write down the first few terms of the weak field
expansions of these functions:
\bear
S_{F12}(z) &=&
G_{F12}\biggl[
1-2{G_{B12}\over T}z^2+{2\over 3}\Bigl({G_{B12}\over T}
+{G_{B12}^2\over T^2}\Bigr)
z^4 +\, {\rm O}(z^6) \biggr]
\, ,
\\
A_{F12}(z) &=&
G_{F12}\dot G_{B12}
\biggl[z-\Bigl(\third + {2\over 3}{G_{B12}\over T}\Bigr)
z^3+\, {\rm O}(z^5)\biggr] \;.
\ear

\item
{\bf The purely magnetic field case:}

\noindent
With the same notations as for the periodic case above, the determinant factor becomes
\bear
{\rm det}^{{1\over 2}}
\bigl[{\rm cos \cZ}
\bigr]
&=&
\cosh z \;.
\label{app-gdpdetextBap}
\ear
The worldline correlators \eqref{app-gd-calGF}, \eqref{app-gd-dcalFGF}
 specialize to 
\begin{eqnarray}
{\cal G}_{F}(\tau_1,\tau_2) &=&G_{F12}\,{g_-} + S_{F12}(z){g_+}
-A_{F12}(z)\, i{r_+}
\, ,
\\
\dot{\cal G}_{F}(\tau_1,\tau_2)
&=&\dot G_{F12}\Eins
-\frac{2z}{T}A_{F12}(z){g_+}
+\frac{2z}{T} S_{F12}(z)\, i{r_+}
\ear
with coincidence limits
\begin{eqnarray}
{\cal G}_F(\tau,\tau) &=& -{\tanh} \, z \,  i{r_+}
\, ,
\\
\dot {\cal G}_F(\tau,\tau) &=& 2\delta(0) \Eins -\frac{2z}{T} {\tanh} \, z \, g_+ \;.
\end{eqnarray}

\item
{\bf The purely electric field case:}

The rules for the transition from the magnetic to the electric field are the same as for the periodic
case above. 

\item
{\bf The crossed-field case:}

As in the periodic case, for a crossed field the determinant factor becomes unity,
and the worldline correlators are given exactly by their weak-field expansion to order
$O(F^2)$, \eqref{app-gd-GFexpand} and \eqref{app-gd-dGFexpand}.

\end{itemize}

\acknowledgments
We would like to thank F. Fillion-Gourdeau and A. Ilderton for helpful
discussions and correspondence.


\end{document}